\documentclass[sigconf]{acmart}

\usepackage{booktabs} 
\usepackage{times}  
\usepackage{helvet}  
\usepackage{courier}  
\usepackage{url}  
\usepackage{graphicx}  
\usepackage{algpseudocode}
\usepackage{float}
\usepackage{algorithm}
\usepackage{mathtools}
\usepackage{amsmath}
\usepackage{caption}
\usepackage{subfigure}
\usepackage{multirow}
\usepackage{ctable}

\copyrightyear{2018}
\acmYear{2018}
\setcopyright{acmcopyright}
\acmConference[CIKM '18]{The 27th ACM International Conference on
	Information and Knowledge Management}{October 22--26, 2018}{Torino,
	Italy}
\acmBooktitle{The 27th ACM International Conference on Information and
	Knowledge Management (CIKM '18), October 22--26, 2018, Torino, Italy}
\acmPrice{15.00}
\acmDOI{10.1145/3269206.3271730}
\acmISBN{978-1-4503-6014-2/18/10}
\newcommand{\squishlist}{
	\begin{list}{$\bullet$}
		{ \setlength{\itemsep}{0pt}
			\setlength{\parsep}{3pt}
			\setlength{\topsep}{3pt}
			\setlength{\partopsep}{0pt}
			\setlength{\leftmargin}{1.5em}
			\setlength{\labelwidth}{1em}
			\setlength{\labelsep}{0.5em} } }
	
	\newcommand{\squishlisttwo}{
		\begin{list}{$\bullet$}
			{ \setlength{\itemsep}{0pt}
				\setlength{\parsep}{0pt}
				\setlength{\topsep}{0pt}
				\setlength{\partopsep}{0pt}
				\setlength{\leftmargin}{2em}
				\setlength{\labelwidth}{1.5em}
				\setlength{\labelsep}{0.5em} } }
		
		\newcommand{\squishend}{
		\end{list}  }


\begin{document}
			\title{Regularizing Matrix Factorization with User and Item Embeddings for Recommendation}


			\author{Thanh Tran, Kyumin Lee}\vspace{0.1in}
			\affiliation{\institution{Worcester Polytechnic Institute, USA}}
			\email{{tdtran, kmlee}@wpi.edu}
			
			
			\author{Yiming Liao, Dongwon Lee}\vspace{0.1in}
			\affiliation{\institution{Penn State University, USA}}
			\email{{yiming, dongwon}@psu.edu}


			\begin{abstract}
				Following recent successes in exploiting both latent factor  and word embedding models in recommendation, we propose a novel \emph{Regularized Multi-Embedding} (RME) based recommendation model that simultaneously encapsulates the following ideas via decomposition: (1) which items a user {\em likes}, (2) which two users {\em co-like} the same items, (3) which two items  users often {\em co-liked}, and (4) which two items  users often {\em co-disliked}.
				In experimental validation, the RME outperforms competing state-of-the-art models in both explicit and implicit feedback datasets, significantly improving Recall@5 by 5.9$\sim$7.0\%, NDCG@20 by 4.3$\sim$5.6\%, and MAP@10 by 7.9$\sim$8.9\%. In addition, under the \emph{cold-start} scenario for users with the lowest number of interactions, against the competing models, the RME outperforms NDCG@5 by 20.2\% and 29.4\% in MovieLens-10M and MovieLens-20M datasets, respectively. Our datasets and source code are available at: {\sf https://github.com/thanhdtran/RME.git}.

			\end{abstract}

			

			
			\maketitle
			
			\section{Introduction}
			\label{sec:introduction}
			
			
			Among popular {\em Collaborative Filtering} (CF) methods in recommendation \cite{SuK09,hu2008collaborative,koren2009matrix,resnick1994grouplens}, in recent years, latent factor models (LFM) using matrix factorization have been widely used. LFM are known to yield relatively high prediction accuracy, are language independent, and allow additional side information to be easily incorporated and decomposed together \cite{agarwal2009regression,wang2011collaborative}. However, most of conventional LFM only exploited positive feedback while neglected negative feedback and treated them as missing data \cite{devooght2015dynamic,hu2008collaborative,pilaszy2010fast,volkovs2015effective}. 
			
			In movie recommender systems, it was observed that many users who enjoyed watching \emph{Thor: The Dark World}, also enjoyed \emph{Thor: Ragnarok}. In this case, \emph{Thor: The Dark World} and \emph{Thor: Ragnarok} can be seen as a pair of co-liked movies. So, if a user preferred \emph{Thor: The Dark World} but never watch \emph{Thor: Ragnarok}, the system can precisely recommend \emph{Thor: Ragnarok} to her (\textbf{first observation}). Similarly, if two users A and B liked the same movies, we can assume A and B have the same movie interests. If user A likes a movie that B has never watched, the system can recommend the movie to B (\textbf{second observation}). In the same manner, we ask if co-occurred disliked movies can provide any meaningful information. We observed that most users, who rated \emph{Pledge This!} poorly (0.8/5.0 on average), also gave a low rating to \emph{Run for Your Wife} (1.3/5.0 on average). If the disliked co-occurrence pattern was exploited, \emph{Run for Your Wife} would not be recommended to other users who did not enjoy \emph{Pledge This!} (\textbf{third observation}). This will help reduce the false positive rate for the recommender systems. The same phenomena would have also occurred in other recommendation domains.
			
			The first two observations are similar to the basic assumptions of item CF and user CF where similar scores between items/users are used to infer the next recommended items for users. Unfortunately, only the first two observations have been exploited in conventional CF. While treating the negative-feedback items differently from missing data led to better results \cite{he2016fast}, to the best of our knowledge, no previous works exploited the \textbf{third observation} to enhance the recommender systems' performance.
			
			\begin{figure}
				\centering
				\includegraphics[width=\linewidth]{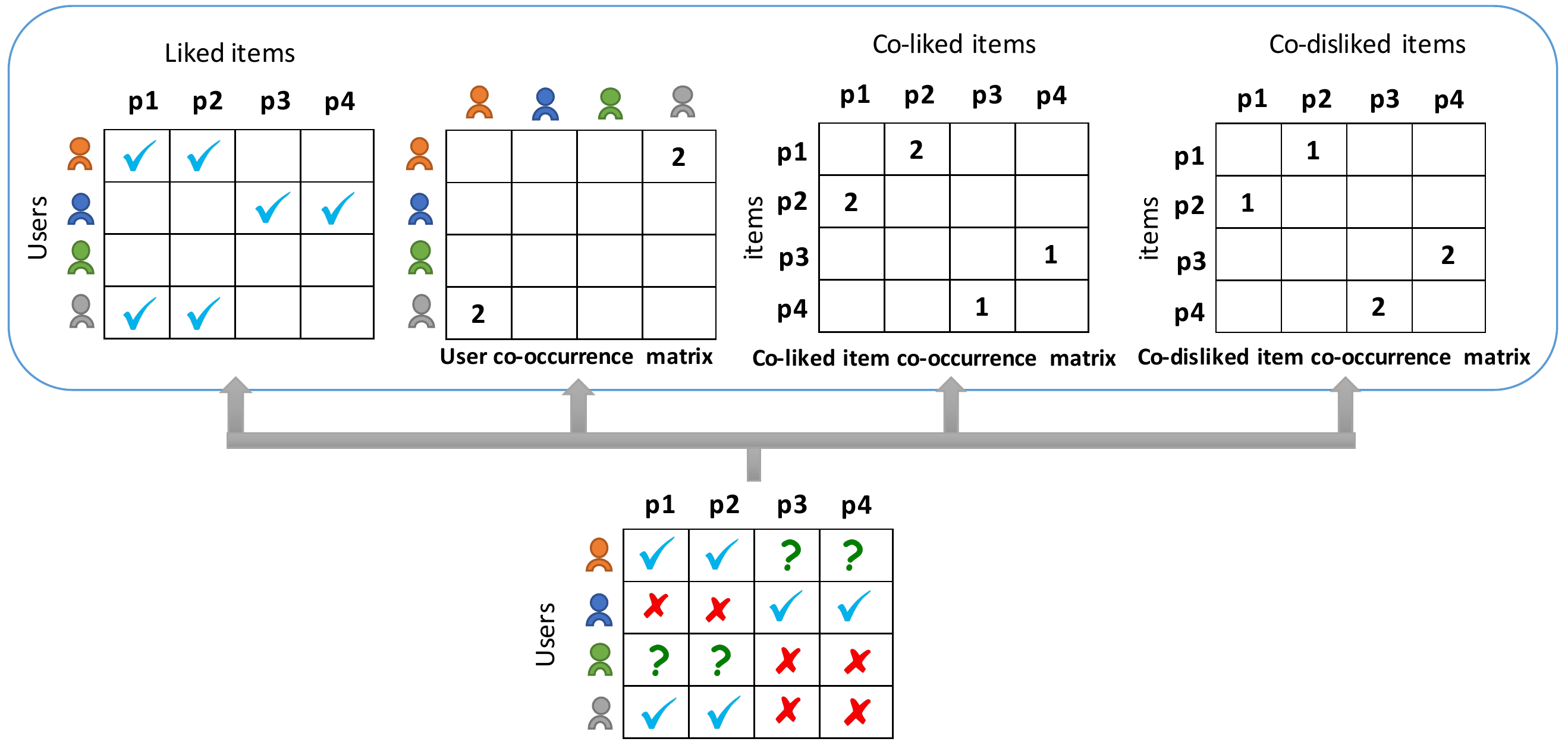}
				\caption{An overview of our RME Model, which jointly decomposes user-item interaction matrix, co-liked item co-occurrence matrix, co-disliked item co-occurrence matrix, and user co-occurrence matrix. ($V$: liked, $X$: disliked, and $?$: unknown)}
				\label{fig:MultiCofactor}
				\vspace{-5pt}
			\end{figure}
			
			Therefore, in this paper, we attempt to exploit all three observations in one model to achieve better recommendation results. With the recent success of word embedding techniques in natural language processing, if we consider pairs of co-occurred liked/disliked items or pairs of co-occurred users as pairs of co-occurred words, we can apply word embedding to learn latent representations of items (e.g., item embeddings) and users (e.g. user embeddings). Based on this, we propose a \emph{Regularized Multi-Embedding} based recommendation model (RME), which jointly decomposes (1) a user-item interaction matrix, (2) a user co-occurrence matrix, (3) a co-liked item co-occurrence matrix, and (4) a co-disliked item co-occurrence matrix. 
			The RME model concurrently exploits the co-liked co-occurrence patterns and co-disliked co-occurrence patterns of items to enrich the items' latent factors. It also augments users' latent factors by incorporating user co-occurrence patterns on their preferred items. Figure~\ref{fig:MultiCofactor} illustrates an overview of our RME model.

			Both liked and disliked items can be explicitly measured by rating scores (e.g., a liked item is $\geq 4$ star-rating and a disliked item is $\leq 2$ star-rating) in explicit feedback datasets such as 5-star rating datasets (e.g., a Movie dataset and an Amazon dataset). However, in implicit feedback datasets (e.g., a music listening dataset and a browsing history dataset), users do not explicitly express their preferences. In \emph{implicit feedback datasets}, the song plays and URL clicks could indicate how much users like the items (i.e., positive samples), but inferring the disliked items (i.e., negative samples) is a big challenge due to the nature of implicit feedback. In order to deal with this challenge, we propose an algorithm which infers a user's disliked items in implicit feedback datasets, so that we can build an RME model and recommend items for both explicit and implicit feedback datasets.
			In this paper, we made the following contributions:
			\squishlist	
			\item We proposed a joint RME model, which combined weighted matrix factorization, co-liked item embedding, co-disliked item embedding, and user embedding, for both explicit and implicit feedback datasets.		
			\item We designed a user-oriented EM-like algorithm to draw negative samples (i.e., disliked items) from implicit feedback datasets.
			\item We conducted comprehensive experiments and showed that the RME model substantially outperformed several baseline models in both explicit and implicit feedback datasets.		
			\squishend

			\section{Preliminaries}
			\label{sec:preliminaries}

			
			\vspace{0.05in}
			\noindent\textbf{Item.} Items are objects that users interact with or consume. They can be interpreted in various ways, depending on the context of a dataset. For example, an item is a movie in a movie dataset such as MovieLens, whereas it is a song in TasteProfile.
			
			\vspace{0.05in}	
			\noindent\textbf{Liked items and disliked items.} 
			In explicit feedback datasets such as MovieLens (a 5-star rating dataset), an item $\geq 4$ stars is classified to a liked item of the user, and an item $\leq 2$ stars is classified to a disliked item of the user \cite{blattner2007exploring}. In implicit feedback datasets such as TasteProfile, the more a user consumes an item, the more he/she likes it (e.g., larger play count in TasteProfile indicates stronger preference). But, disliked items are not explicitly observable.
			
			\vspace{0.05in}
			\noindent\textbf{Top-N recommendation.} In this paper, we focus on top-N recommendation scenario, in which a recommendation model suggests a list of top-N most appealing items to users. We represent the interactions between users and items by a matrix $M^{m*n}$ where \emph{m} is the number of users and \emph{n} is the number of items. If a user $u$ likes an item $p$, $M_{up}$ will be set to 1. From \emph{M}, we are interested in extracting co-occurrence patterns including liked item co-occurrences, disliked item co-occurrences, and user co-occurrences. Our goal is to exploit those co-occurrence information to learn the latent representations of users and items, then recommend top-N items to the users.

			\vspace{0.05in}
			\noindent\textbf{Notations.} Table~\ref{table:Notations} shows key notations used in this paper. Note that all vectors in the paper are column vectors.
			\begin{table}
				\centering
				\small
				\caption{Notations.}
				\vspace{-10pt}
				\label{table:Notations}
				\resizebox{\linewidth}{!}{
					\begin{tabular}{c|l}
						\toprule
						Notation 	& Description \\
						\midrule
						$M$ & a $m \times n$ user-item interaction matrix. \\
						$U$ & a $m \times k$ latent factor matrix of users. \\
						$P$ & a $n \times k$ latent factor matrix of items. \\
						$X$& a $n \times n$ SPPMI matrix of liked items-item co-occurrences.\\
						$Y$ & a $n \times n$ SPPMI matrix of disliked item-item co-occurrences.\\
						$Z$ & a $m \times m$ SPPMI matrix of user-user co-occurrences. \\
						$\alpha_u$ & a $k \times 1$ latent factor vector of user $u$. \\
						$\beta_p$ & a $k \times 1$ latent factor vector of item $p$. \\
						$\gamma_i$ & a $k \times 1$ latent factor vector of co-liked item context $i$. \\
						$\delta_{i\prime}$ & a $k \times 1$ latent factor vector of co-disliked item context $i\prime$.\\
						$\theta_j$ & a $k \times 1$ latent factor vector of user context $j$. \\
						$\lambda $ & a hyperparameter of regularization terms. \\
						$b, d$ & co-liked and co-disliked item bias. \\
						$c, e$ & co-liked and co-disliked item context bias. \\
						$f, g$ & user bias and user context bias. \\
						$w_{up}$ & a weight for an interaction between user $u$ and her liked item $p$. \\
						$w_{uj}^{(u)}$ & a weight for two users $u$ and $j$ who co-liked same items.\\
						$w_{pi}^{(+p)}$ & a weight for two items $p$ and $i$ that are co-liked by users. \\
						$w_{pi}^{(-p)}$ & a weight for two items $p$ and $i$ that are co-disliked by users. \\
						\bottomrule
					\end{tabular}
				}
				\vspace{-15pt}
			\end{table}	
			
			\section{Our RME Model}
			\label{sec:rec-models}
			First, we review the Weighted Matrix Factorization (WMF), and co-liked item embedding. Then, we propose co-disliked item embedding and user embedding. Finally, we describe our RME model and present how to compute it.
			
			\subsection{WMF, Embedding and RME model}
			\noindent\textbf{Weighted matrix factorization (WMF).}
			WMF is a widely-used collaborative filtering method in recommender systems \cite{hu2008collaborative}. Given a sparse user-item matrix $M^{m \times n}$, the basic idea of WMF is to decompose \emph{M} into a product of 2 low rank matrices $U^{m \times k}$ and $P^{n \times k}$ (i.e., $M = U  \times  P ^T$ ), where $k$ is the number of dimensions and $k < min(m,n)$. Here, $U$ is interpreted as a latent factor matrix of users, and $P$ is interpreted as a latent factor matrix of items.

			We denote $U^T = ({\alpha_1, \alpha_2, ..., \alpha_m})$ where $\alpha_u \in R^k$ ($u \in \overline{1, m}$) and $\alpha_u$ represents the latent factor vector of user $u$. Similarly, we denote $P^T = ({\beta_1, \beta_2, ...,\beta_n})$ where $\beta_p \in R^k$ ($p \in \overline{1, n}$) and $\beta_p$ represents the latent factor vector of item $p$. The objective of WMF is defined by:
			\begin{equation}
			\label{equa:wmf}
			\noindent
			\resizebox{0.435\textwidth}{!}{$
				\begin{aligned}
				\mathcal{L}_{WMF} &= \frac{1}{2}\sum_{u,p}^{}w_{up}(M_{up} - \alpha^T_u\beta_p)^2 +
				\frac{1}{2}\bigg( \lambda_\alpha\sum_{u}^{}||\alpha_u||^2 + \lambda_\beta\sum_{p}^{}||\beta_p||^2 \bigg)
				\end{aligned}
				$}
			\end{equation}
			where $w_{up}$ is a hyperparameter to compensate the interaction between user $u$ and item $p$, and is used to balance between the number of non-zero and zero values in a sparse user-item matrix. The weight $w$ of the interaction between user $u$ and item $p$ (denoted as $w_{up}$) can be set as $w_{up} = l(1 + \phi M_{up})$ \cite{hu2008collaborative,liang2016factorization} where $l$ is a relative scale and $\phi$ is a constant.
			$\lambda_\alpha$ and $\lambda_\beta$ are used to adjust the importance of two quadratic regularization terms $\sum_{u}^{}||\alpha_u||^2$ and $\sum_{p}^{}||\beta_p||^2$. 

			\vspace{0.05in}
			\noindent\textbf{Word embedding models.} Word embedding models have recently received a lot of attention from the research community. 
			Given a sequence of training words, the embedding models learn a latent representation for each word. For example, \emph{word2vec} \cite{mikolov2013distributed} is one of popular word embedding methods. Especially, the skip-gram model in word2vec tries to predict surrounding words (i.e., word context) of a given word in the training set.

			According to Levy et al. \cite{levy2014neural}, skip-gram model with negative sampling (SGNS) is equivalent to implicitly factorize a word-context matrix, whose cells are the \emph{Pointwise Mutual Information} (PMI) of the respective word and context pairs, shifted by a global constant. Let $D$ as a collection of observed word and context pairs, the PMI between a word $i$ and its word context $j$ is calculated as:
			\begin{equation}
			\nonumber
			PMI(i, j) = log{\frac{P(i,j)}{P(i)*P(j)}}
			\vspace{-5pt}
			\end{equation}
			where $P(i,j)$ is the joint probability that word $i$ and word $j$ appears together within a window size (e.g. $P(i,j) = \frac{\#(i,j)}{|D|}$, where $|D|$ refers to the total number of word and word context pairs in $D$). Similarly, $P(i)$ is the probability the word $i$ appears in $D$, and $P(j)$ is the probability word $j$ appears in $D$ (e.g. $P(i) =\frac{\#(i)}{|D|}$ and $P(j) =\frac{\#(j)}{|D|}$). Obviously, $PMI(i,j)$ can be calculated as:
			\begin{equation}
			\label{equa:pmi}
			PMI(i, j) = log{\frac{\#(i,j)*|D|}{\#(i)*\#(j)}}
			\vspace{-5pt}
			\end{equation}
			By calculating $PMI$ of all word-context pairs in $D$, we can form a squared $n\times n$ matrix $M^{PMI}$ where $n$ is the total number of distinct words in $D$. Next, a \emph{Shifted Positive Pointwise Mutual Information} (SPPMI) of two words $i$ and $j$ is calculated as:
			\begin{equation}
			\label{equa:sppmi}
			SPPMI(i,j) = max(PMI(i, j) - log(s), 0)
			\vspace{-5pt}
			\end{equation}
			where $s$ is a hyperparameter to control the density of PMI matrix $M^{PMI}$ and $s$ can be interpreted equivalently as a hyperparameter that indicates the number of negative samples in SGNS. When $s$ is large, more values in the matrix $M^{PMI}$ are cleared, leading $M^{PMI}$ to become sparser. When $s$ is small, matrix $M^{PMI}$ becomes denser.  Finally, factorizing matrix $M^{SPPMI}$, where each cell in $M^{SPPMI}$ is transformed by Formula (\ref{equa:sppmi}), is equivalent to performing SGSN.
			
			\vspace{0.05in}
			\noindent\textbf{Co-liked item embedding (LIE).} As mentioned in the previous studies \cite{guardia2015latent,liang2016factorization,barkan2016item2vec}, when users liked/consumed items in a sequence, the items sorted by the ascending interaction time order can be inferred as a sequence. Thus, performing co-liked item embeddings to learn latent representations of items is equivalent to perform word embeddings to learn latent representations of words. Therefore, we can apply word embedding methods to learn latent representations of items, and perform a joint learning between embedding models and traditional factorization methods (e.g. WMF).
			
			Given each user's liked item list, we generate co-liked item-item co-occurrence pairs without considering liked time. Particularly, given a certain item in the item sequence, we consider all other items as its contexts. We call this method as a \emph{greedy context generation} method which can be applied to other non-timestamped datasets. After generating item and item context pairs, we construct an item co-occurrence SPPMI matrix and perform SPPMI matrix factorization. In particular, given generated item-item co-occurrence pairs, we construct a SPPMI matrix of items by applying Equation (\ref{equa:pmi}) to calculate the pointwise mutual information of each pair, and then by measuring the shifted positive pointwise mutual information of the pair based on Equation (\ref{equa:sppmi}). Once the SPPMI matrix of co-liked items is constructed, we incorporate it to the traditional matrix factorization method to improve the item latent representations.

			\vspace{0.05in}
			\noindent\textbf{Co-disliked item embedding (DIE).} 
			As mentioned in the Introduction section, when many users disliked two items $p_1$ and $p_2$ together, the two items can form a pair of co-occurred disliked items. If the recommender systems learned this disliked co-occurrence pattern, it would not recommend item $p_2$ to a user, who disliked $p_1$. This will help reduce the false positive rate for the recommender systems. Therefore, similar to liked item embeddings, we applied the word embedding technique to exploit the disliked co-occurrence information to enhance the item's latent factors.
			
			\vspace{0.05in}
			\noindent\textbf{User embedding (UE).}  When two users A and B preferred same items, we can assume the two users share similar interests. Therefore, if user A enjoyed an item $p$ that has not been observed in user B's transactions, we can recommend the item to user B. Similar to liked and disliked item embeddings, we applied the word embedding technique to learn user embeddings that explain the co-occurrence patterns among users.
			
			From the user-item interaction matrix $M^{m \times n}$, where each row represents consumed items of a user (e.g. a list of items that the user rated or backed), we only keep liked items per user in the matrix $M'$. Then, we construct a $n \times m$ reverse matrix $M'^T$ of $M'$, where each row represents users that liked a certain item. Then, users, who liked the same item, form a sequence, and the sequence of users is interpreted as a sequence of words. From this point, word embedding techniques are applied to the user sequence to enhance latent representations of users. 
			
			\vspace{0.05in}
			\noindent\textbf{Our RME model.} It is a joint learning model combining WMF, co-liked item embedding, co-disliked item embedding, and user embedding. It minimizes the following objective function:
			
			\vspace{-10pt}
			\begin{equation}
			\label{func:obj-model5}
			\resizebox{0.9\linewidth}{!}{$
				\begin{aligned}
				\mathcal{L} &= 	
				\overbrace{\frac{1}{2}\sum_{u,p}^{}w_{up}(M_{up} - \alpha^T_u\beta_p)^2}^{\mathcal{L}_1} \text{  (WMF)}  \\
				&+	\overbrace{\frac{1}{2}\sum_{X_{pi} \ne 0}^{}  w^{(+p)}_{pi}   (X_{pi} - \beta^T_p\gamma_i - b_p - c_i)^2}^{\mathcal{L}_2}  \text{  (LIE)} \\
				&+	\overbrace{\frac{1}{2}\sum_{Y_{p{i\prime}} \ne 0}^{}  w^{(-p)}_{p{i\prime}}  (Y_{p{i\prime}} - \beta^T_p\delta_{i\prime} - d_p - e_{i\prime})^2}^{\mathcal{L}_3}  \text{  (DIE)} \\
				&+	\overbrace{\frac{1}{2}\sum_{Z_{uj} \ne 0}^{}  w^{(u)}_{uj}  (Z_{uj} - \alpha^T_u\theta^{}_j - f^{}_u - g^{}_j)^2}^{\mathcal{L}_4} \text{  (UE)} \\
				&+	\frac{1}{2}\lambda
				\bigg(
				\sum_{u}^{}||\alpha_u||^2
				+  			\sum_{p}^{}||\beta_p||^2 +
				\sum_{i}^{}||\gamma_i||^2 +
				\sum_{{i\prime}}^{}||\delta_{i\prime}||^2
				+			\sum_{j}^{}||\theta^{}_j||^2
				\bigg)
				\end{aligned}
				$}
			\vspace{-5pt}
			\end{equation}
			where the item's latent representation $\beta_p$ is shared among WMF, co-liked item embedding and co-disliked item embedding. The user's latent representation $\alpha_u$ is shared between WMF and user embedding. $X$ and $Y$ are SPPMI matrices, constructed by co-liked item-item co-occurrence patterns and disliked item-item co-occurrence patterns, respectively. $\gamma$ and $\delta$ are $k \times 1$ latent representation vectors of co-liked item context and co-disliked item context, respectively. $Z$ is a SPPMI matrix constructed by user-user co-occurrence patterns. $\theta$ is a $k \times 1$ latent representation vector of a user context. $w^{(+p)}$, $w^{(-p)}$ and $w^{(u)}$ are hyperparameters to compensate for item/user co-occurrences in $X$, $Y$ and $Z$ when performing decomposition. $b$ is liked item bias, and $c$ is co-liked item context bias. $d$ is disliked item bias, and $e$ is co-disliked item-context bias. $f$ and $g$ are user bias and user context bias, respectively. Incorporating bias terms were originally introduced in \cite{Koren09}. A liked item bias $b_p$ and a co-liked item context bias $c_i$ mean that when the two items $p_i$ and $p_j$ are co-liked by users, each item may have a little bit higher/lower preference compared to the average preference. The similar explanation is applied to the other biases. The last line show regularization terms along with a hyperparameter $\lambda$ to control their effects.

			\subsection{Optimization}
			
			We can use the stochastic gradient descent to optimize the Equation (\ref{func:obj-model5}). However, it is not stable and sensitive to parameters \cite{yu2014parallel}. Therefore, we adopt vector-wise ALS algorithm \cite{zhou2008large,yu2014parallel} that alternatively optimize each model's parameter in parallel while fixing the other parameters until the model gets converged. Specifically, we calculate the partial derivatives of the model's objective function with regard to the model parameters (i.e., \{$\alpha_{1:m}$, $\beta_{1:n}$, $\gamma_{1:n}$, $\delta_{1:n}$, $b_{1:n}$, $c_{1:n}$, $d_{1:n}$, $e_{1:n}$, $\theta_{1:m}$, $f_{1:m}$, $g_{1:m}$\}). Then we set them to zero and obtain updating rules. Details are given as follows:
			
			From the objective function in Equation (\ref{func:obj-model5}), while taking partial derivatives of $\mathcal{L}$ with regard to each user's latent representation vector $\alpha_u$, we observe that only $\mathcal{L}_1$, $\mathcal{L}_4$ and the L2 user regularization $\frac{1}{2}\lambda \sum_{u}^{}||\alpha_u||^2$ contain $\alpha_u$. Therefore, we obtain:
			
			\vspace{-10pt}
			\begin{equation}
			\nonumber
			\resizebox{0.9\linewidth}{!}{$
				\begin{aligned}
				\frac{\partial \mathcal{L}}{\partial \alpha_u} &= \frac{\partial \mathcal{L}_1}{\partial \alpha_u} + \frac{\partial \mathcal{L}_4}{\partial \alpha_u} + \frac{\partial \lambda \sum_{u}^{}||\alpha_u||^2}{2\partial \alpha_u} \\
				&=-\sum_{u,p}^{}w_{up}(M_{up} - \alpha^T_u\beta_p)\beta_p^T
				-\sum_{u,j}^{}  w^{(u)}_{uj}  (Z^{}_{uj} - \alpha^T_u\theta^{}_j - f^{}_u - g^{}_j)\theta_j^T + \lambda \alpha_u^T
				\end{aligned}
				$}
			\end{equation}
			
			\noindent Fixing item latent vectors $\beta$, user context latent vectors $\theta$, user bias $d$ and user context bias $e$, and solving $\frac{\partial \mathcal{L}}{\partial \alpha_u} = 0$, we obtain the updating rule of $\alpha_u$ as follows:
			
			\vspace{-10pt}
			\begin{equation}
			\label{func:update-alpha}
			\resizebox{0.7\linewidth}{!}{$
				\begin{aligned}
				\alpha_u =&
				\bigg[
				\sum_{p}^{}w_{up}\beta_p\beta_p^T +
				\sum_{j|Z_{uj}\ne 0}^{}    w^{(u)}_{uj}    \theta_j\theta_j^T + \lambda I_K
				\bigg]^{-1}  \\
				& \bigg[
				\sum_{p}^{}w_{up}M_{up}\beta_p +
				\sum_{j|Z_{uj} \ne 0}^{}    w^{(u)}_{uj}    (Z_{uj} - f_u - g_j)\theta_j
				\bigg] \\
				& \text{ ,} \forall 1 \le u \le m, 1 \le p \le n, 1 \le j \le m
				\end{aligned}
				$}
			\end{equation}
			
			Similarly, taking partial derivatives of $\mathcal{L}$ with respect to each item latent vector $\beta_p$ needs to consider only $\mathcal{L}_1$, $\mathcal{L}_2$, $\mathcal{L}_3$ and item regularization $\frac{1}{2}\lambda \sum_{p}^{}||\beta_p||^2$. By fixing other parameters and solving $\frac{\partial \mathcal{L}}{\partial \beta_p} = 0$, we obtain:
			
			\vspace{-10pt}
			\begin{equation}
			\label{func:update-beta}
			\resizebox{0.9\linewidth}{!}{$
				\begin{aligned}
				\beta_p  =
				& \bigg[
				\sum_{u}^{}w_{up}\alpha_u\alpha_u^T  +
				\sum_{i|X_{pi} \ne 0}^{}     w^{(+p)}_{pi}    \gamma_i \gamma^{T}_i  +
				\sum_{i\prime|Y_{pi\prime} \ne 0}^{}  w^{(-p)}_{pi\prime}    \delta_{i\prime}\delta^{T}_{i\prime}  +
				\lambda I_K
				\bigg] ^ {-1} \\
				&	
				\bigg[
				\sum_{u}^{}w_{up}M_{up}\alpha_u +
				\sum_{i|X_{pi} \ne 0}^{}     w^{(+p)}_{pi}    (X_{pi} - b_p - c_i) \gamma_i  +\\
				&	\sum_{i\prime|Y_{pi\prime} \ne 0}^{}      w^{(-p)}_{pi\prime}    (Y_{pi\prime} - d_p - e_{i\prime})  \delta_{i\prime}
				\bigg] \\
				& \text{ ,} \forall 1 \le u \le m, 1 \le p \le n, 1 \le i, i^{\prime} \le n
				\end{aligned}
				$}
			\end{equation}
			
			In the same manner, we obtain the update rules of item contexts $\gamma$, $\delta$, and user context $\theta$ alternatively as follows:
			\vspace{-5pt}
			\begin{equation}
			\label{func:update-context}
			\resizebox{0.81\linewidth}{!}{$
				\begin{aligned}
				\gamma_i =&
				\bigg[
				\sum_{p|X_{ip} \ne 0}^{}    w^{(+p)}_{ip}    \beta_p\beta_p^T + \lambda I_K
				\bigg]^{-1} 		
				\bigg[
				\sum_{p|X_{ip} \ne 0}^{}    w^{(+p)}_{ip}    (X_{ip} - b_p - c_i)\beta_p
				\bigg]
				\\
				\delta_{i\prime} =&
				\bigg[
				\sum_{p|Y_{{i\prime}p} \ne 0}^{}    w^{(-p)}_{{i\prime}p}    \beta_p\beta_p^T +  \lambda I_K
				\bigg]^{-1}
				\bigg[
				\sum_{p|Y_{{i\prime}p} \ne 0}^{}    w^{(-p)}_{{i\prime}p}    (Y_{{i\prime}p} - d_p - e_{i\prime})\beta_p
				\bigg]
				\\
				\theta_j =&  		
				\bigg[
				\sum_{u|Z_{ju} \ne 0}^{}    w^{(u)}_{ju}    \alpha_u\alpha_u^T + \lambda I_K
				\bigg]^{-1}
				\bigg[
				\sum_{u|Z_{ju} \ne 0}^{}    w^{(u)}_{ju}    (Z_{ju} - d_u - e_j)\alpha_u
				\bigg] \\				
				& \text{ ,} \forall 1 \le u \le m, 1 \le p \le n, 1 \le i, i^{\prime} \le n
				\end{aligned}
				$}
			\end{equation}
			
			
			The item biases and item context biases $b$, $c$, $d$, $e$, as well as the user and user context biases $f$, $g$ are updated alternatively using the following update rules:
			\begin{equation}
			\label{func:update-biases}
			\resizebox{0.65\linewidth}{!}{$
				\begin{aligned}
				b_p =& \frac{1}{|i:X_{pi} \ne 0|} \sum_{i:X_{pi} \ne 0}^{} (X_{pi} - \beta_p^T\gamma_i - c_i)
				\\
				c_i =& \frac{1}{|p:X_{ip} \ne 0|} \sum_{p:X_{ip} \ne 0}^{} (X_{ip} - \beta_p^T\gamma_i - b_p)
				\\
				d_p =& \frac{1}{|i\prime:Y_{p i\prime} \ne 0|} \sum_{i\prime:Y_{pi\prime} \ne 0}^{} (Y_{pi\prime} - \beta_p^T\delta_{i\prime} - e_{i\prime})
				\\
				e_{i\prime} =& \frac{1}{|p:Y_{i\prime p} \ne 0|} \sum_{p:Y_{i\prime p} \ne 0}^{} (Y_{i\prime p} - \beta_p^T\delta_{i\prime} - d_p)
				\\
				f_u &= \frac{1}{|j:Z_{uj} \ne 0|} \sum_{j:Z_{uj} \ne 0}^{} (Z_{uj} - \alpha_u^T\theta_j - g_j)
				\\
				g_j &= \frac{1}{|u:Z_{ju} \ne 0|} \sum_{u:Z_{ju} \ne 0}^{} (Z_{ju} - \alpha_u^T\theta_j - f_u)
				\end{aligned}
				$}
			\end{equation}
			
			In short, the pseudocode of our proposed RME model is presented in Algorithm \ref{alg:mcf-algorithm}.
			\begin{algorithm}
				\caption{RME algorithm}	
				\begin{algorithmic}[1]
					
					\small			
					\Require~~ M, $\lambda$
					\State Build SPPMI matrices of liked item $X$, disliked item $Y$ and user co-occurrences $Z$ using Eq. (\ref{equa:pmi}) and Eq. (\ref{equa:sppmi})
					\State Initialize $U$ (or $\alpha_{1:m}$), $P$ (or $\beta_{1:n}$), $\gamma_{1:n}, \delta_{1:n}, \theta_{1:m}$.
					\State Initialize $b_{1:n}, c_{1:n}, d_{1:n}, e_{1:n}, f_{1:m}, g_{1:m}$.
					\Repeat
					\State For each user u, update $\alpha_u$ by Eq. (\ref{func:update-alpha}) ($1 \le u \le m$).
					\State For each item p, update $\beta_p$ by Eq. (\ref{func:update-beta}) ($1 \le p \le n$).
					
					\State Alternatively update each item context $\gamma_i$, $\delta_{i^{\prime}}$ and user context $\theta_{j}$  by Eq. (\ref{func:update-context}) ($1 \le i, i^{\prime} \le n$; $1 \le j \le m$).
					\State Alternatively update each bias $b_{p}, c_{i}, d_{p}, e_{i^{\prime}}, f_u, g_j$ by Eq. (\ref{func:update-biases}) ($1 \le p, i, i^{\prime} \le n$; $1 \le u, j \le m$).
					\Until convergence
					\State \Return $U, P$
					
				\end{algorithmic}
				\label{alg:mcf-algorithm}
			\end{algorithm}
			\vspace{-10pt}

			\subsection{Complexity Analysis}
			In this section, we briefly provide time complexity analysis of our model. Let ${\Omega}_M$ = \{(u, p) | $M_{up} \ne 0$\}
			, $\Omega_X$ = \{(p, i) | $X_{pi} \ne 0$ \}, $\Omega_Y$ = \{(p, $i^{\prime}$) | $Y_{pi^{\prime}}  \ne 0$\}, $\Omega_Z$ = \{(u, $j$) | $Z_{uj} \ne 0$\}. Constructing SPPMI matrices X, Y and Z take $O(|\Omega_X|^2)$, $O(|\Omega_Y|^2)$ and $O(|\Omega_Z|^2)$, respectively. However, the SPPMI matrices are calculated once and are constructed in parallel using batch processing, so they are not costly. For learning RME model, computing $\alpha$ takes $O((|\Omega_M| + |\Omega_Z|)k^2 + k^3)$ time, and computing $\beta$ takes $O((|\Omega_M| + |\Omega_X| + |\Omega_Y|)k^2 + k^3)$ time. Also, it takes $O(|\Omega_X|k^2 + k^3)$ for computing co-liked item context $\gamma$, and so do other latent contexts $\delta$, $\theta$. It takes $O(|\Omega_Z|k)$ time to compute all user bias $f$ and so do the other biases. Thus, the time complexity for RME is $O(\eta(2(|\Omega_M| + |\Omega_X| + |\Omega_Y| + |\Omega_Z|)k^2 + (2m + 3n)k^3))$, where $\eta$ is the number of iterations. Since $k << min(m,n)$ and M, X, Y, Z are often sparse, which mean ($|\Omega_M| + |\Omega_X| + |\Omega_Y| + |\Omega_Z|$) is small, the time complexity of RME is shortened as $O(\eta(m + \frac{3}{2}n)k^3)$, which scales linearly to the conventional ALS algorithm for collaborative filtering \cite{yu2014parallel}.

			\section{Inferring Disliked Items in Implicit Feedback datasets}
			Unlike  explicit feedback datasets, there is a lack of substantial evidence, on which items the users disliked in implicit feedback datasets. Since our model exploits co-disliked item co-occurrences patterns among items, the implicit feedback datasets challenge our model. To deal with this problem, we can simply assume that missing values are equally likely to be negative feedback, then sample some negative instances from missing values with uniform weights \cite{volkovs2015effective,steck2010training,pilaszy2010fast,he2017neural}. However, assigning uniform weight is suboptimal because the missing values are a mixture of negative and unknown feedbacks. A recent work suggests to sample negative instances by assigning non-uniform weights based on item popularity \cite{he2016fast}. The idea is that popular items are highly aware by users, so if they are not observed in a user's transactions, it assumes that the user dislikes them. However, this sampling method is also not optimal because same unobserved popular items can be sampled across multiple users. This approach does not reflect each user's personalized interest. 
			
			Instead, we follow the previous works \cite{zhang2013optimizing,pan2008one,liu2002partially}, and propose a user-oriented EM-like algorithm to draw negative samples (i.e., inferred disliked items) for users in implicit feedback datasets. Our approach is described as follows:
			
			First, we assume that an item with a low ranking score of being liked will have a higher probability to be drawn as a negative sample of a user. Given $r_u$ is the ranked list of all items of the user $u$, the prior probabilities of items to be drawn as negative samples are calculated by using a softmax function as follows:
			\begin{equation}
			\label{func:manified-softmax}
			Pr^{(u)}_i = \frac{\exp{(-r_u[i])}} {\sum_{j=1}^{n} \exp{(-r_u[j])}}
			\end{equation}

			After negative samples are drawn for each user, we built the RME model by using Algorithm \ref{alg:mcf-algorithm}. The pseudocode of the RME model for implicit feedback datasets is presented in Algorithm \ref{alg:mcf-implicit}.
			
			In Algorithm \ref{alg:mcf-implicit}, since each user may prefer a different number of items, we define a hyper-parameter $\tau$ as a negative sample drawing ratio to control how many negative samples we will sample for each user. In line 6, $count(u)$ returns the number of observed items of a user $u$. Then, the number of drawn negative samples for the user $u$ is calculated and assigned to $ns$. If a user prefers 10 items and $\tau=0.8$, the algorithm will sample 8 disliked items. We note that sampling with replacement is used such that different items are drawn independently. The value of $\tau$ is selected using the validation data. In line 8, we set the ranking of observed items to $+\infty$ to avoid drawing the observed items as negative samples. In line 12, we build the RME model based on the negative samples drawn in the Expectation step, and temporally store newly learned user latent matrix, item latent matrix and corresponding NDCG to $U\_tmp, P\_tmp, ndcg$ variables, respectively (NDCG is a measure to evaluate recommender systems, which will be mentioned in Experiment section). If we obtain a better $ndcg$ comparing with the previous NDCG $prev\_ndcg$ (line 13), we will update $U, P, prev\_ndcg$ with new values (line 14). Overall, at the end of the Expectation step, we obtain the disliked items for each user. Then, in the Maximization step, we build our RME model to re-learn user and item latent representations $U$ and $P$. The process is repeated until getting converged or the early stopping condition (line 13 to 17) is satisfied. 
			
			\begin{algorithm}[t]	
				\caption{RME model for implicit feedback datasets using user-oriented EM-like algorithm to draw negative samples}
				\begin{algorithmic}[1]
					\Require~~ M, negative sample drawing ratio $\tau$
					\State $max\_iter$ = 10, $prev\_ndcg$ = 0, $iter$ = 0
					\State Initialize Step: $U, P = WMF(M)$
					\Repeat
					\State $iter$ += 1
					\State \Comment{Expectation Step}
					\For{$u \in [1,m]$}:
					\State $ns$ = $\tau$ * $count$(u)
					\State Compute ranked item list: $r_u = {P}.\alpha_u$
					\State Assign observed items with ranking of $+\infty$.
					\State Measure prior probabilities of items to be drawn as negative samples by Eq. (\ref{func:manified-softmax}) then randomly draw $ns$ negative samples with those prior probabilities.
					\EndFor
					\State \Comment{Maximization Step with early stopping}
					\State $U\_tmp, P\_tmp, ndcg$ = RME(train\_data, vad\_data)
					\If {$ndcg > prev\_ndcg$}
					\State $U, P, prev\_ndcg = U\_tmp, P\_tmp, ndcg$
					\Else
					\State break \Comment{Early stopping}
					\EndIf
					\Until $iter$ < $max\_iter$
					\State \Return $U, P$
				\end{algorithmic}
				\label{alg:mcf-implicit}
			\end{algorithm}
			
			\vspace{0.02in}
			\noindent\textbf{Time Complexity:} In order to construct RME model for implicit feedback datasets, we need to re-learn RME model, which includes re-building 3 SPPMI matrices in the maximization step in $\eta\prime$ iterations to get converged. Thus, it takes $O(\eta\prime( (|\Omega_X|^2 + |\Omega_Y|^2 + |\Omega_Z|^2) + \eta(m + \frac{3}{2}n)k^3))$ time where $\eta\prime$ is small.

			\begin{table*}[t]
				\centering
				\small
				\caption{Performance of the baselines, our RME model, and its two variants. The improvement of our model over the baselines and its variants were significant with \emph{p-value} $<$ 0.05 in the three datasets under the non-directional two-sample t-test.}
				\vspace{-5pt}
				\label{table:PerformanceComparison}
				\resizebox{0.9\textwidth}{!}{
					\begin{tabular}{ccccccccccc}
						\toprule
						\multirow{2}{*}{\text{Method}} & \multicolumn{3}{c}{\textbf{MovieLens-10M}} &\multicolumn{3}{c}{\textbf{MovieLens-20M}} & \multicolumn{3}{c}{\textbf{TasteProfile}} 
						\\
						& Recall@5 & NDCG@20 & MAP@10 & Recall@5 & NDCG@20 & MAP@10 &Recall@5 & NDCG@20 & MAP@10
						\\
						\midrule
						Item-KNN 	& 0.0137 & 0.0338 & 0.0397 & 0.0131 & 0.0345 & 0.0402 & 0.0793 & 0.0685 & 0.0904 \\
						Item2vec 	& 0.1020 & 0.1001 & 0.0502 & 0.1066 & 0.1019 & 0.0539 & 0.1455 & 0.1593 & 0.0727 \\
						WMF 		& 0.1280 & 0.1245 & 0.0655 & 0.1348 & 0.1290 & 0.0720 & 0.1745 & 0.1853 & 0.0931 \\
						Cofactor 	& 0.1460 & 0.1381 & 0.0772 & 0.1480 & 0.1387 & 0.0804 & 0.1771 & 0.1873 & 0.0950 \\
						\midrule
						U\_RME 		& 0.1516 & 0.1412 & 0.0818 & 0.1524 & 0.1425 & 0.0847 & 0.1825 & 0.1899 & 0.0997 \\
						I\_RME 		& 0.1511 & 0.1422 & 0.0817 & 0.1530 & 0.1412 & 0.0838 & 0.1826 & 0.1915 & 0.0996 \\
						RME 		& \textbf{0.1562} & \textbf{0.1458} & \textbf{0.0841} & \textbf{0.1570} & \textbf{0.1461} & \textbf{0.0869} & \textbf{0.1876} & \textbf{0.1954} & \textbf{0.1025} \\
						\bottomrule
					\end{tabular}
				}
			\end{table*}

			\section{Experiments}
			\label{sec:exp}

			\subsection{Experimental Settings}
			
			\noindent\textbf{Datasets:} 	
			To measure the performance of our RME model, we evaluate the model on 3 real-world datasets:
			\squishlist	
			\item MovieLens-10M \cite{resnick1994grouplens}: is an explicit feedback dataset. It consists of 69,878 users and 10,677 movies with 10m ratings. Following the k-cores preprocessing \cite{he2016ups,he2017neural}, we only kept users, who rated at least 5 movies, and movies, which were rated by at least 5 users. This led to 58,057 users and 7,223 items (density$=0.978\%$). 
			\item MovieLens-20M: is an explicit feedback dataset. It consists of 138,000 users, 27,000 movies, and 20 millions of ratings. We filtered with the same condition as for MovieLens-10M. This led to 111,146 users and 9,888 items (density$=0.745\%$). 
			\item TasteProfile: is an implicit feedback dataset containing a song's play count by a user \footnote{http://the.echonest.com/}. The play counts are user's implicit preference and are binarized. Similar to the preprocessing at \cite{liang2016factorization}, we first subsampled the dataset to 250k users and 25k items. Then we kept only users, who listened to at least 20 songs, and songs, which were listened by at least 50 users. As a result, 221,011 users and 22,713 songs were remained (density$=0.291$\%).
			\squishend	
			
			\noindent\textbf{Baselines:} To illustrate the effectiveness of our RME model, we compare it with the following baselines:
			\squishlist
			\item WMF \cite{hu2008collaborative}: It is a weighted matrix factorization with \emph{l2}-norm regularization.
			\item Item-KNN \cite{deshpande2004item}: This is an item neighborhood-based collaborative filtering method.
			\item Item2Vec \cite{barkan2016item2vec}: This method used Skip-gram with negative sampling \cite{mikolov2013distributed} to learn item embeddings, then adopted a similarity score between item embeddings to generate user's recommendation lists.
			\item Cofactor \cite{liang2016factorization}: This is a method that combines WMF and co-liked item embedding.
			\squishend
			We note that we do not compare our models with user collaborative filtering method (i.e. User-KNN) because it is not applicable to run the method on the large datasets. However, \cite{sarwar2001item} reported that User-KNN had worse performance than Item-KNN, especially when there are many items but few ratings in a dataset.
			
			\noindent\textbf{Our models:} We not only compare the baselines with our RME, but also two variants of our model such as U\_RME and I\_RME to show the effectiveness of incorporating all of the user embeddings, liked-item embeddings and disliked-item embeddings:
			\squishlist
			\item U\_RME (i.e., RME - DIE): This is a variant of our model, considering only WMF, user embeddings, and liked-item embeddings.
			\item I\_RME (i.e., RME - UE): This is another variant of our model, considering only WMF, liked-item embeddings, and disliked-item embeddings.
			\item RME: This is our proposed RME model.
			\squishend

			\vspace{0.02in}
			\noindent\textbf{Evaluation metrics.} We used three well-known ranking-based metrics -- Recall@N, normalized discounted cumulative gain (NDCG@N), and mean average precision (MAP@N). Recall@N considers all items in top $N$ items equally, whereas NDCG@N and MAP@N apply an increasing discount of $log_2$ to items at lower ranks.
			
			\vspace{0.02in}
			\noindent\textbf{Training, validation and test sets.} Some researchers adopted leave-one-out evaluation \cite{he2017neural,xue2017deep}, but it is not a realistic scenario \cite{he2016fast}. Therefore, we follow 70/10/20 proportions for splitting the original dataset into training/validation/test sets \cite{liang2016modeling}. MovieLens-10M and MovieLens-20M datasets contain timestamp values of user-movie interactions. To create training/validation/testing sets for these datasets, we sorted all user-item interaction pairs in the ascending interaction time order in each of MovieLens-10M and MovieLens-20M datasets. The first 80\% was used for training and validation, and the rest 20\% data was used as a test set. 
			Out of 80\% data extracted for training and validation, we randomly took 10\% for the validation set. To measure the statistical significance of RME over the baselines, we repeated the splitting process five times (i.e., generating five pairs of training and validation sets). Since TasteProfile dataset did not contain timestamp information of user-song interactions, we randomly split the TasteProfile dataset into training/validation/test sets five times with 70/10/20 proportions. Averaged results are reported in the following subsection.
			
			\begin{figure*}
				\centering
				\subfigure[NDCG@N on MovieLens-10M. ] 
				{
					\label{fig:ml10m-vary-N}
					\includegraphics[width=0.22\textwidth]{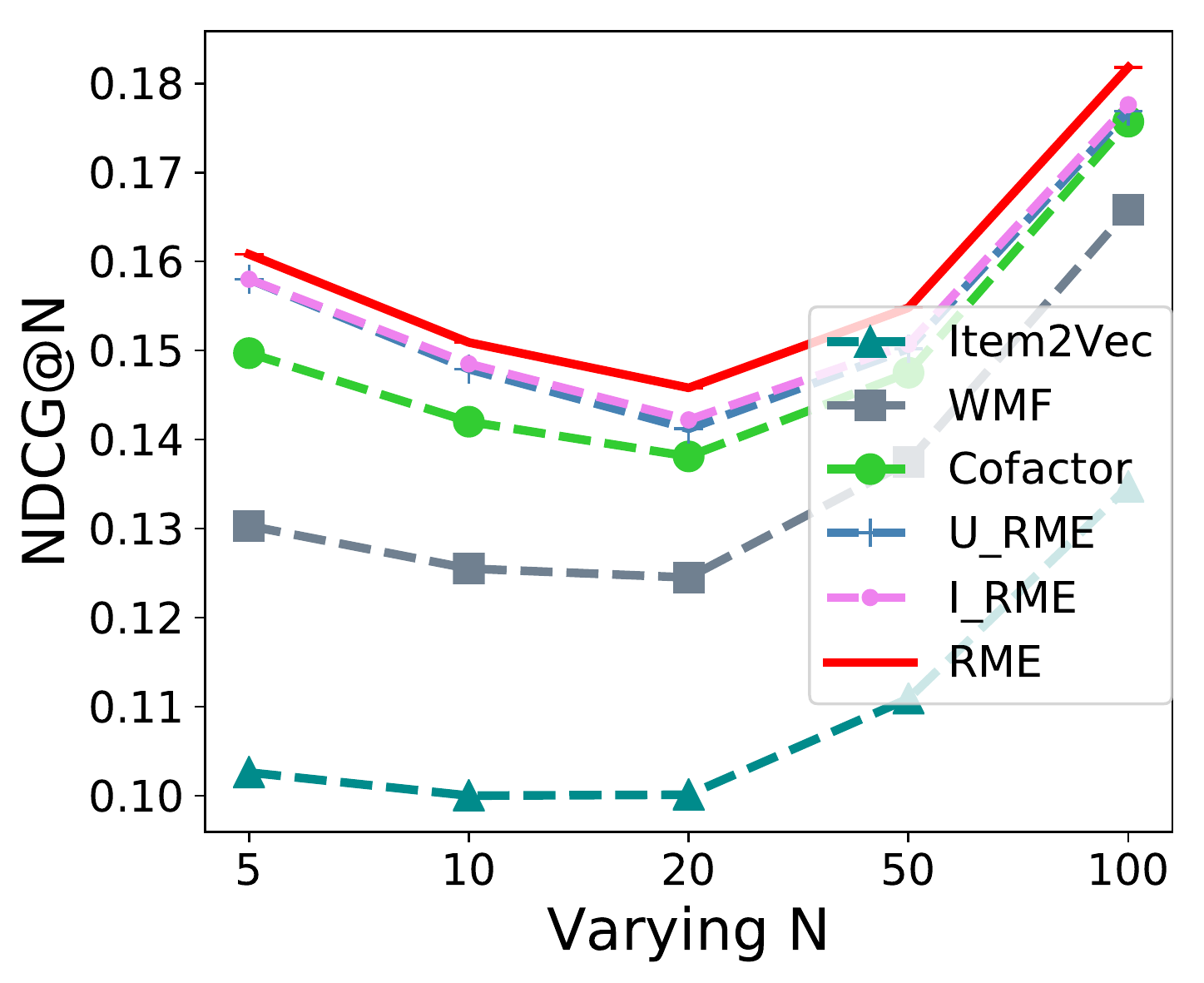}
				}\vspace{-5pt}
				\subfigure[NDCG@N on MovieLens-20M. ] 
				{
					\label{fig:ml20m-vary-N}
					\includegraphics[width=0.22\textwidth]{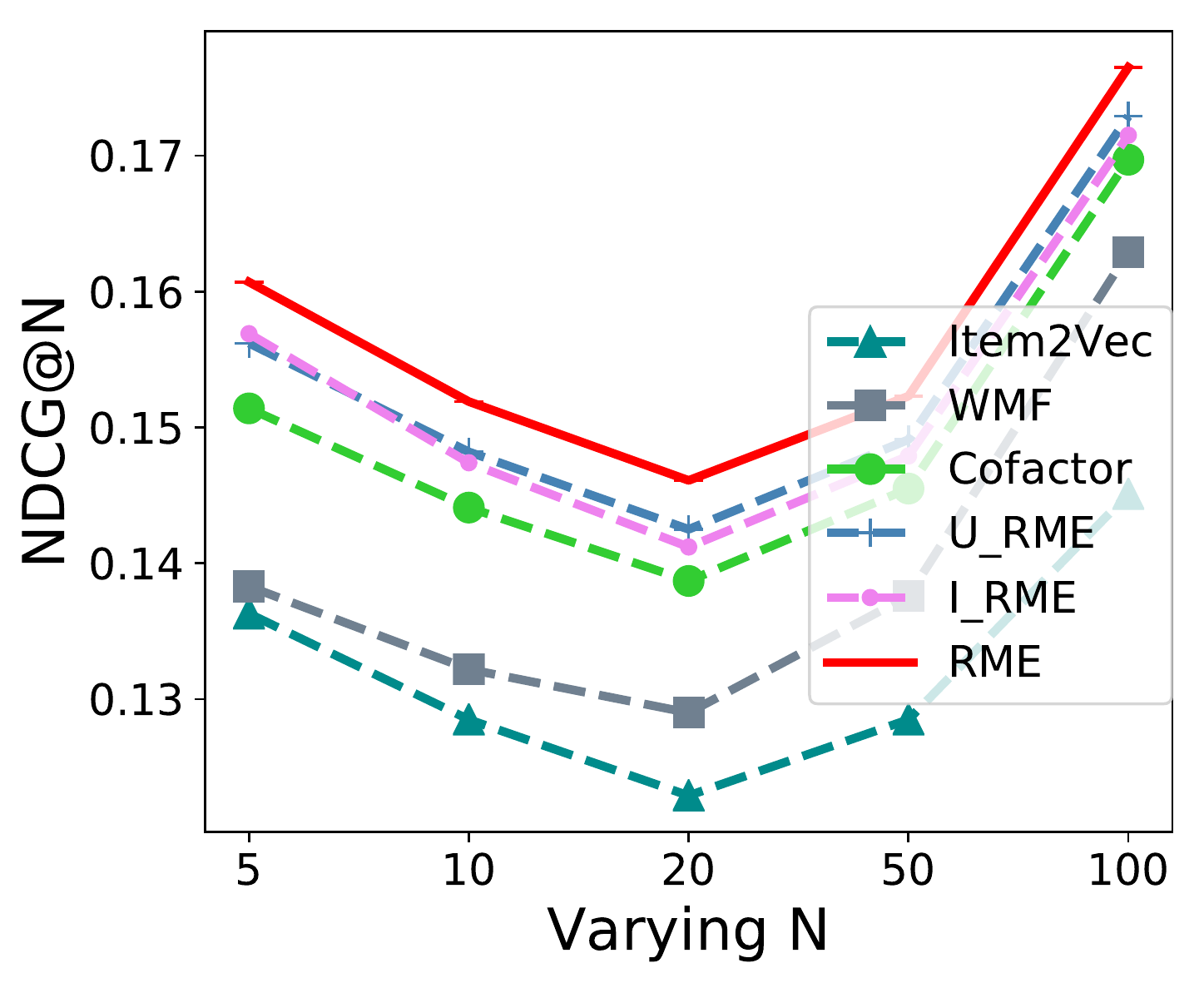}
				}\vspace{-5pt}
				\subfigure[NDCG@N on TasteProfile. ] 
				{
					\label{fig:tp-vary-N}
					\includegraphics[width=0.22\textwidth]{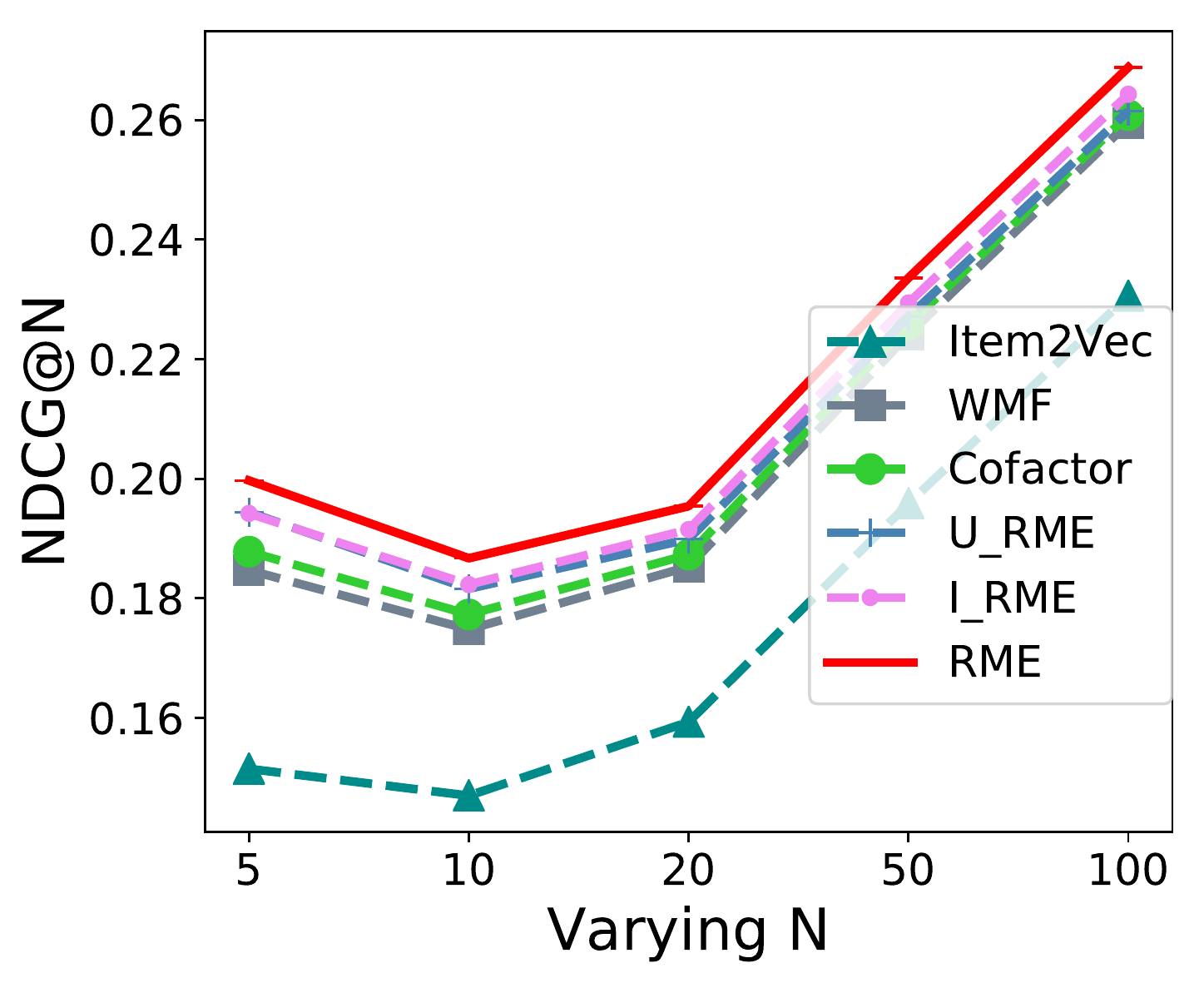}
				}
				\caption{Performance of all models when varying top $N$.}
				\label{fig:vary-N}
				\vspace{-10pt}
			\end{figure*}
			
			\vspace{0.02in}
			\noindent\textbf{Stopping criteria and Hyperparameters.} To decide when to stop training a model, we measured the model's $NDCG@100$ by using the validation set. We stopped training the model when there was no further improvement. Then, we applied the best model to the test set to evaluate its performance. This method was applied to the baselines and RME.

			All hyper-parameters were tuned on the validation set by a grid search. We used the same hyper-parameter setting in all models. The grid search of the regularization weight $\lambda$ was performed in \{0.001, 0.005, 0.01, 0.05, ..., 10\}. The size of latent dimensions was in a range of \{30, 40, 50, ..., 100\}. We set weights $w^{(+p)} = w^{(-p)} = w^{(u)} = w$ for all user-user and item-item co-occurrence pairs. When building our RME model for TasteProfile dataset, we do a grid search for the negative sample drawing ratio $\tau$ in \{0.2, 0.4, 0.6, 0.8, 1.0\}. 

			\subsection{Experimental Results}

			\begin{figure}
				\centering
				\subfigure[Recall@5, NDCG@5, MAP@5 on MovieLens-10M. Fix $\lambda=1$, and vary $k$.] 
				{
					\label{fig:ml10m-k}
					\includegraphics[width=0.155\textwidth]{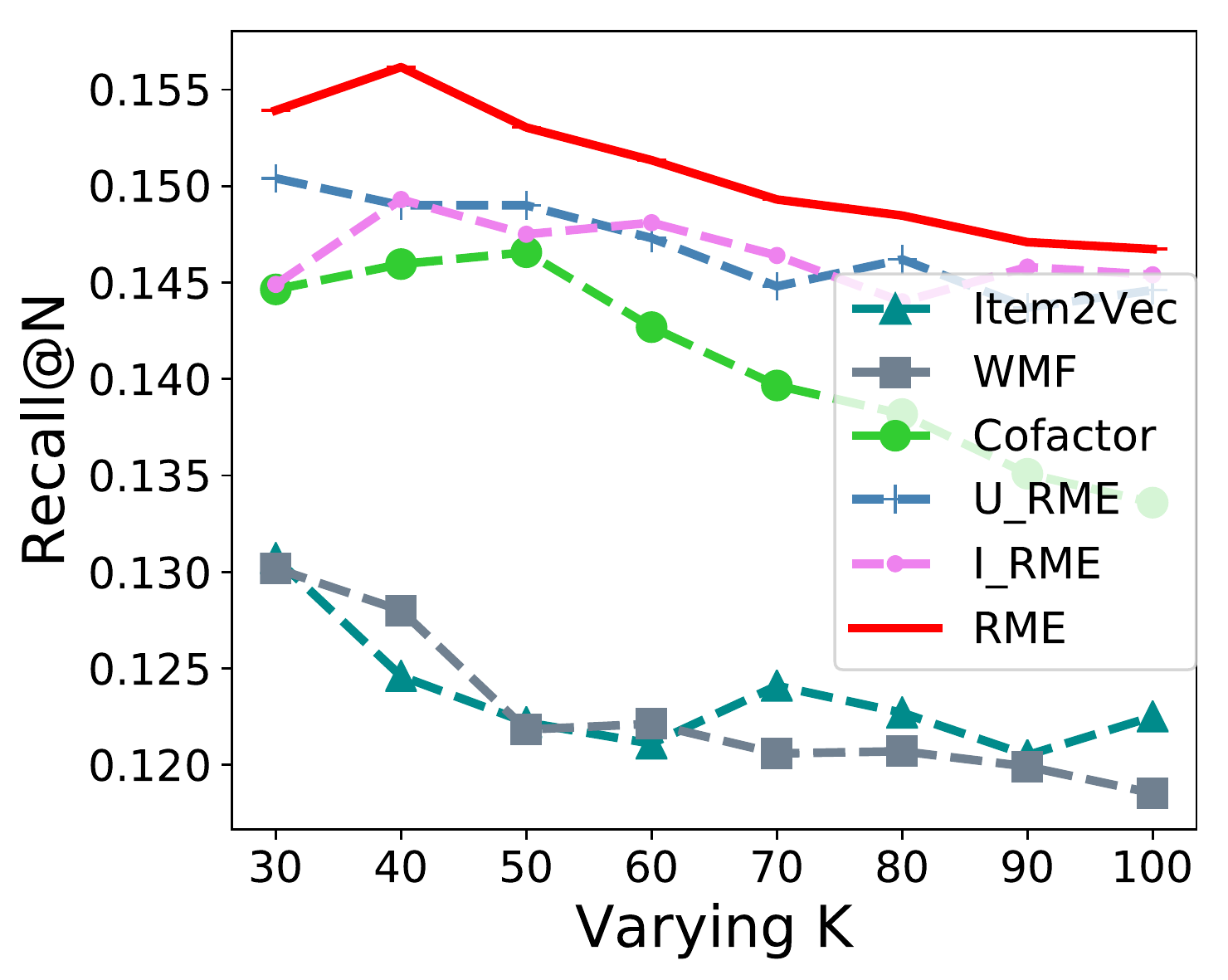}
					\includegraphics[width=0.155\textwidth]{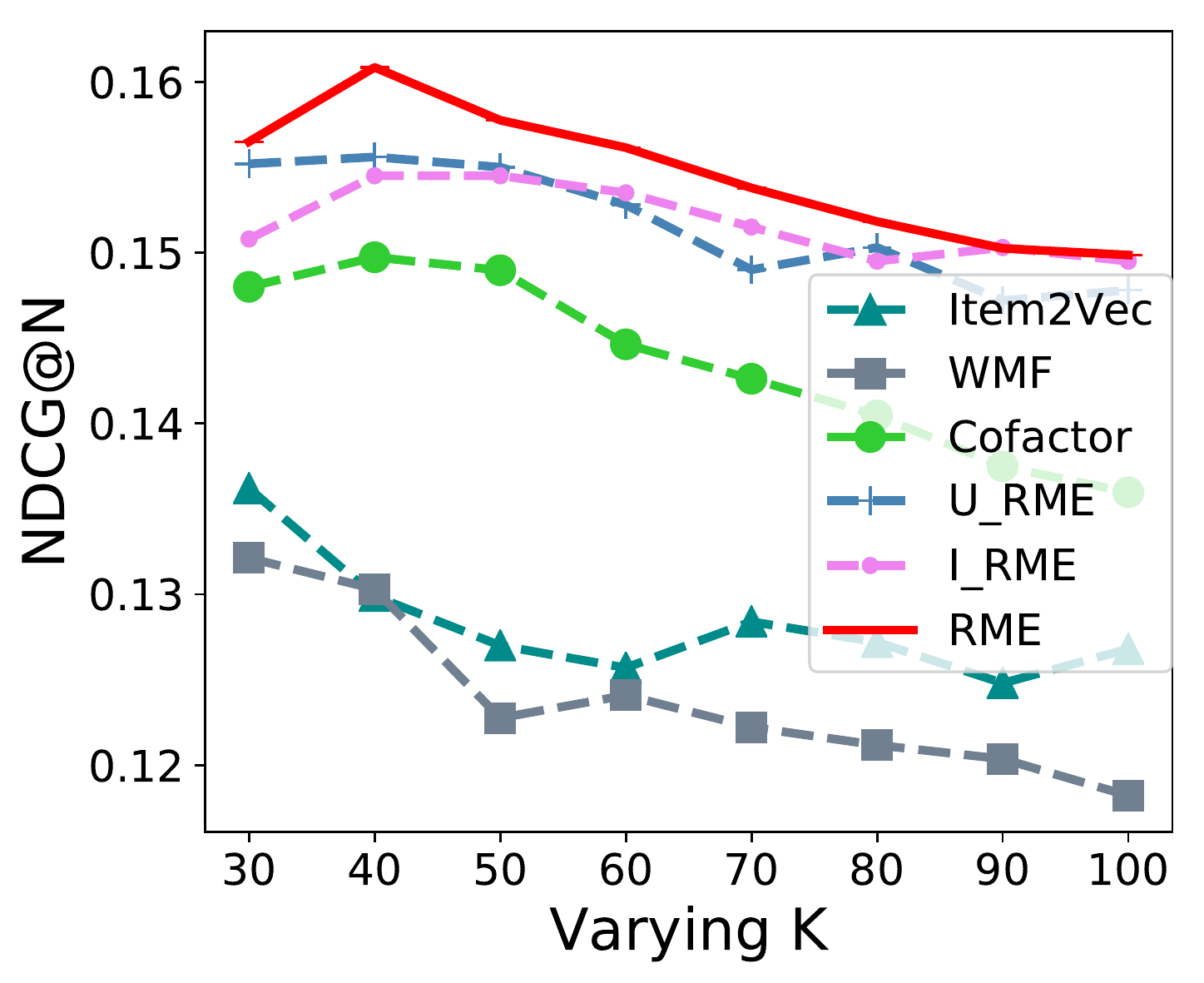}
					\includegraphics[width=0.155\textwidth]{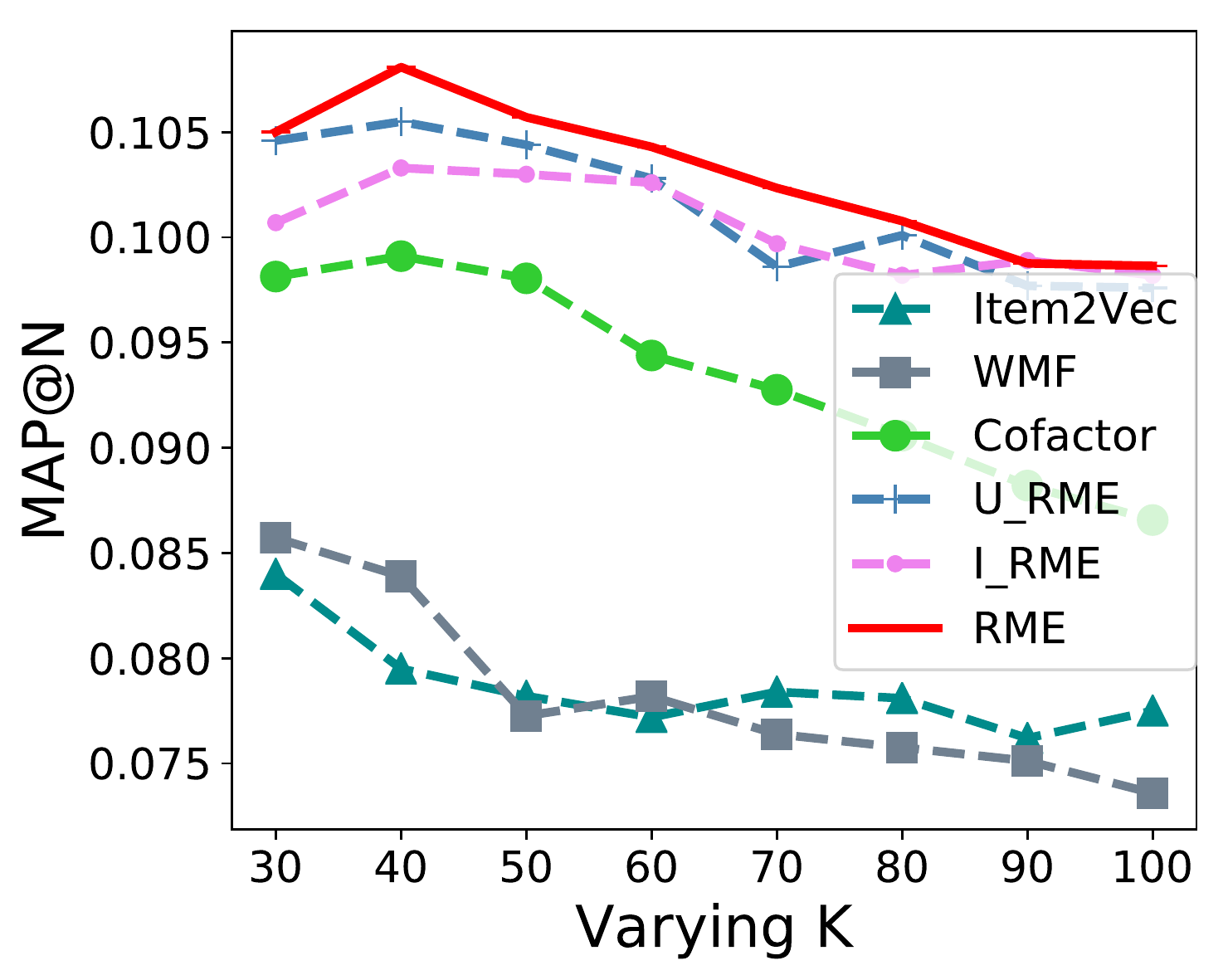}
				}\vspace{-5pt}
				\subfigure[Recall@5, NDCG@5, MAP@5 on MovieLens-20M. Fix $\lambda = 0.5$, and vary $k$.] 
				{
					\label{fig:ml20m-k}
					\includegraphics[width=0.155\textwidth]{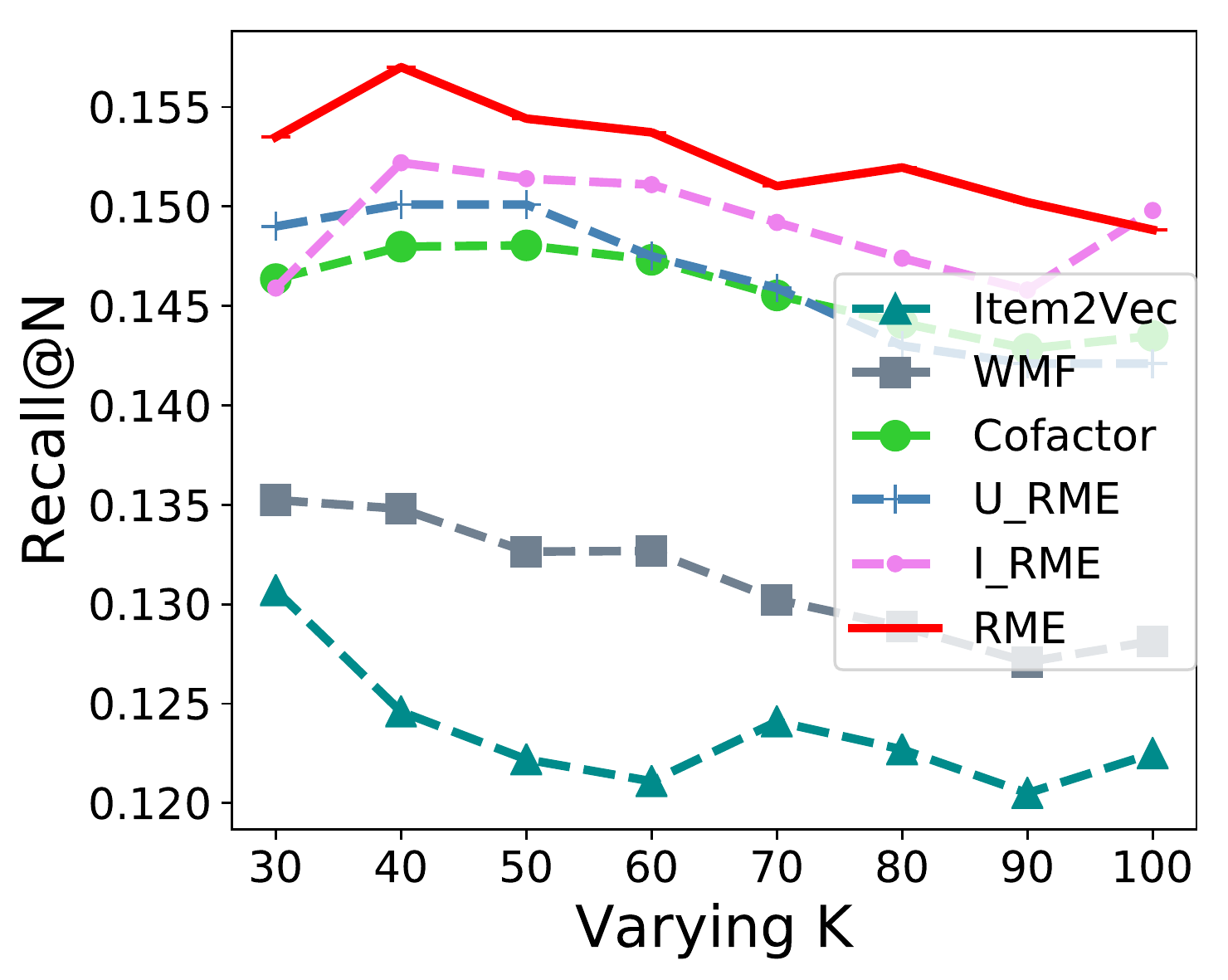}
					\includegraphics[width=0.155\textwidth]{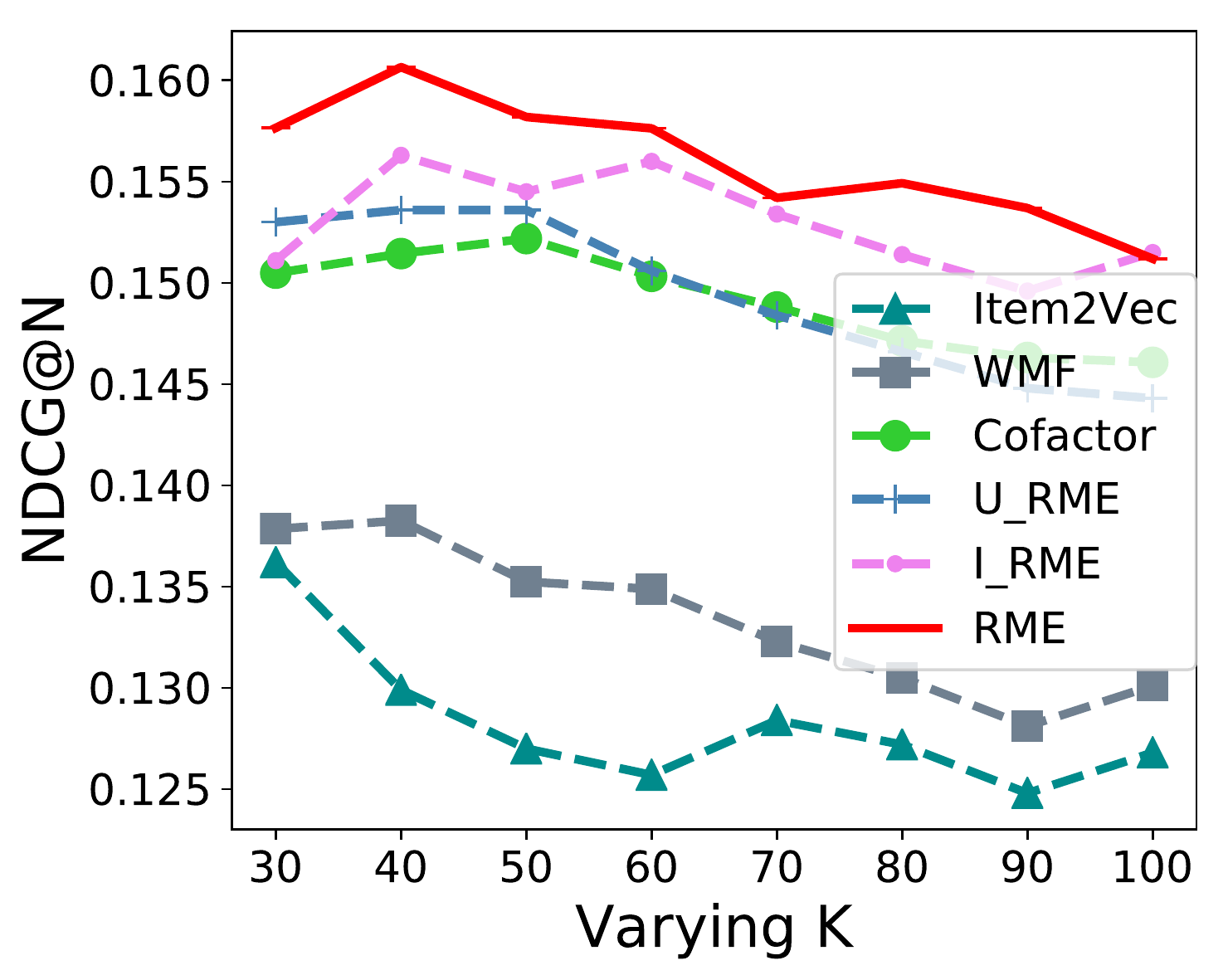}
					\includegraphics[width=0.155\textwidth]{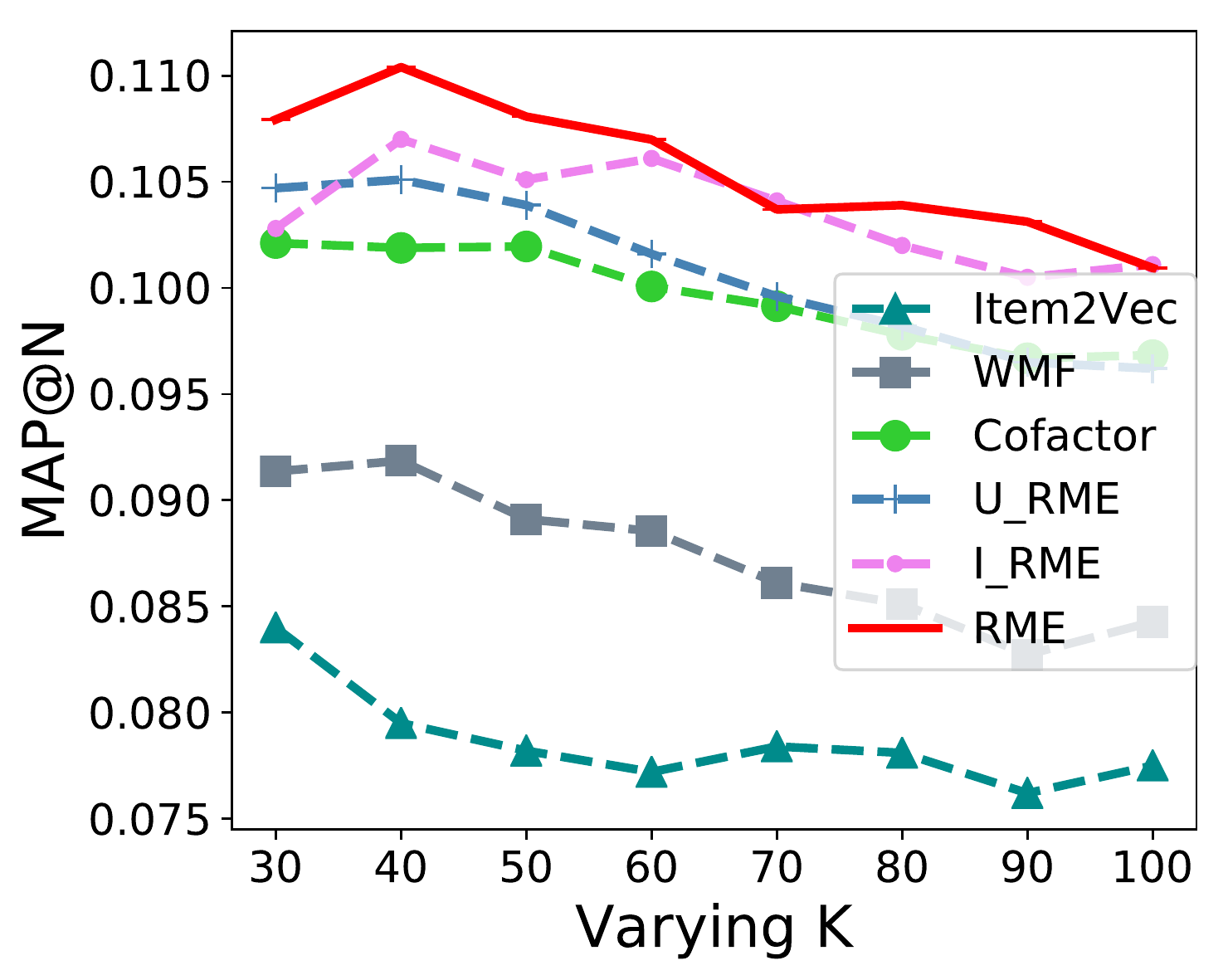}
				}\vspace{-5pt}
				\subfigure[Recall@5, NDCG@5, MAP@5 on TasteProfile. Fix $\lambda = 10$, $\tau=0.2$, and vary $k$.] 
				{
					\label{fig:tp-k}
					\includegraphics[width=0.155\textwidth]{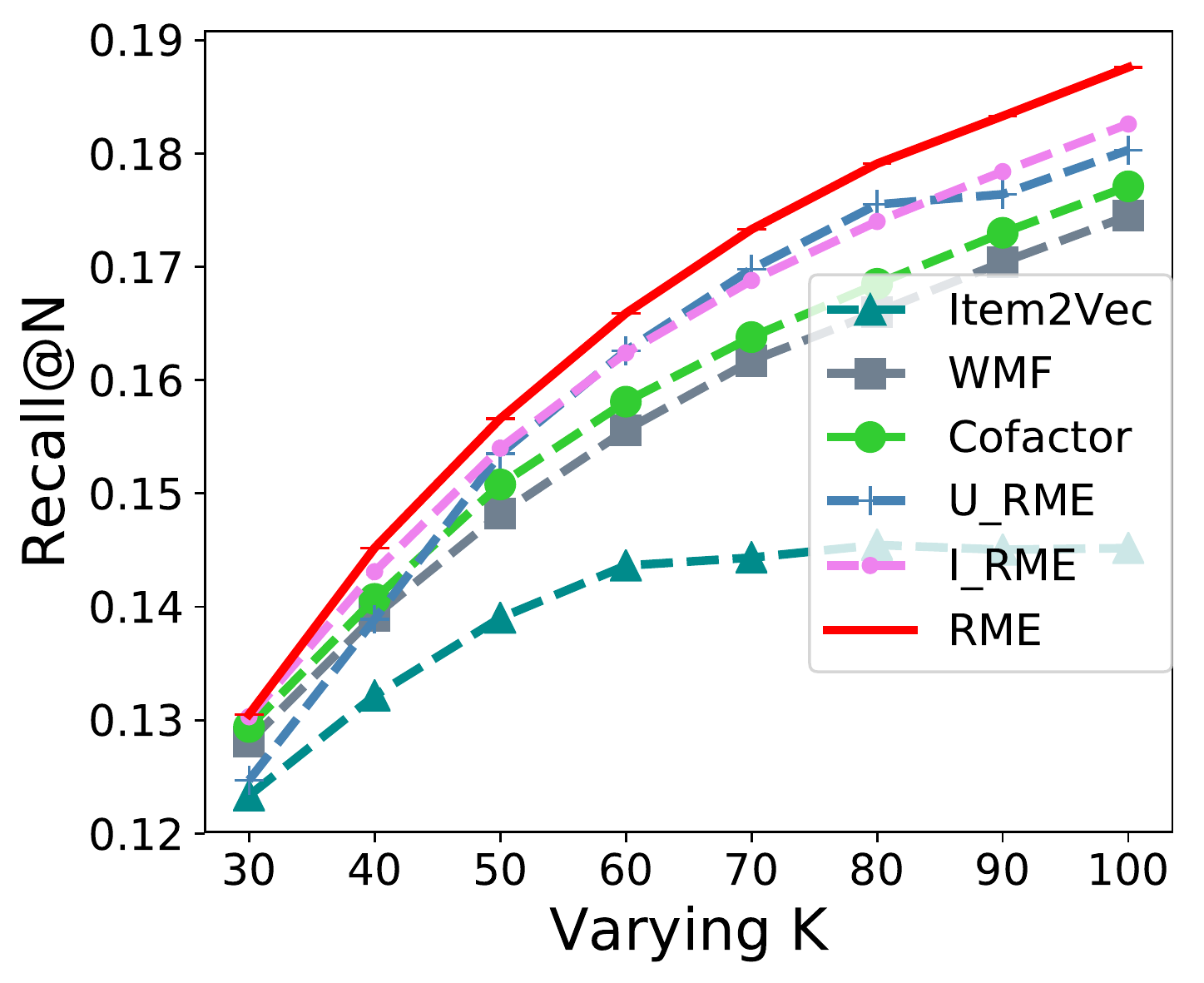}
					\includegraphics[width=0.155\textwidth]{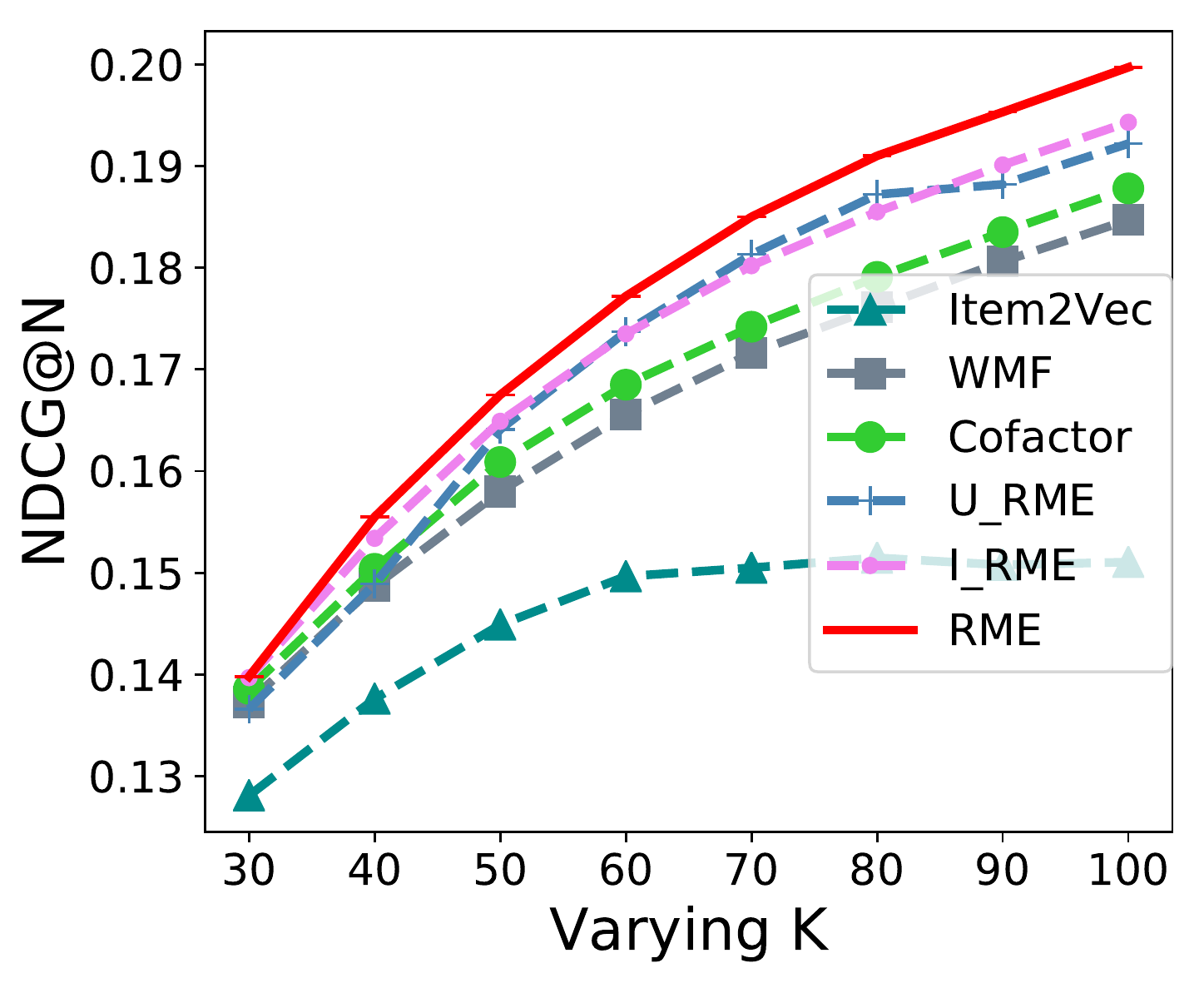}
					\includegraphics[width=0.155\textwidth]{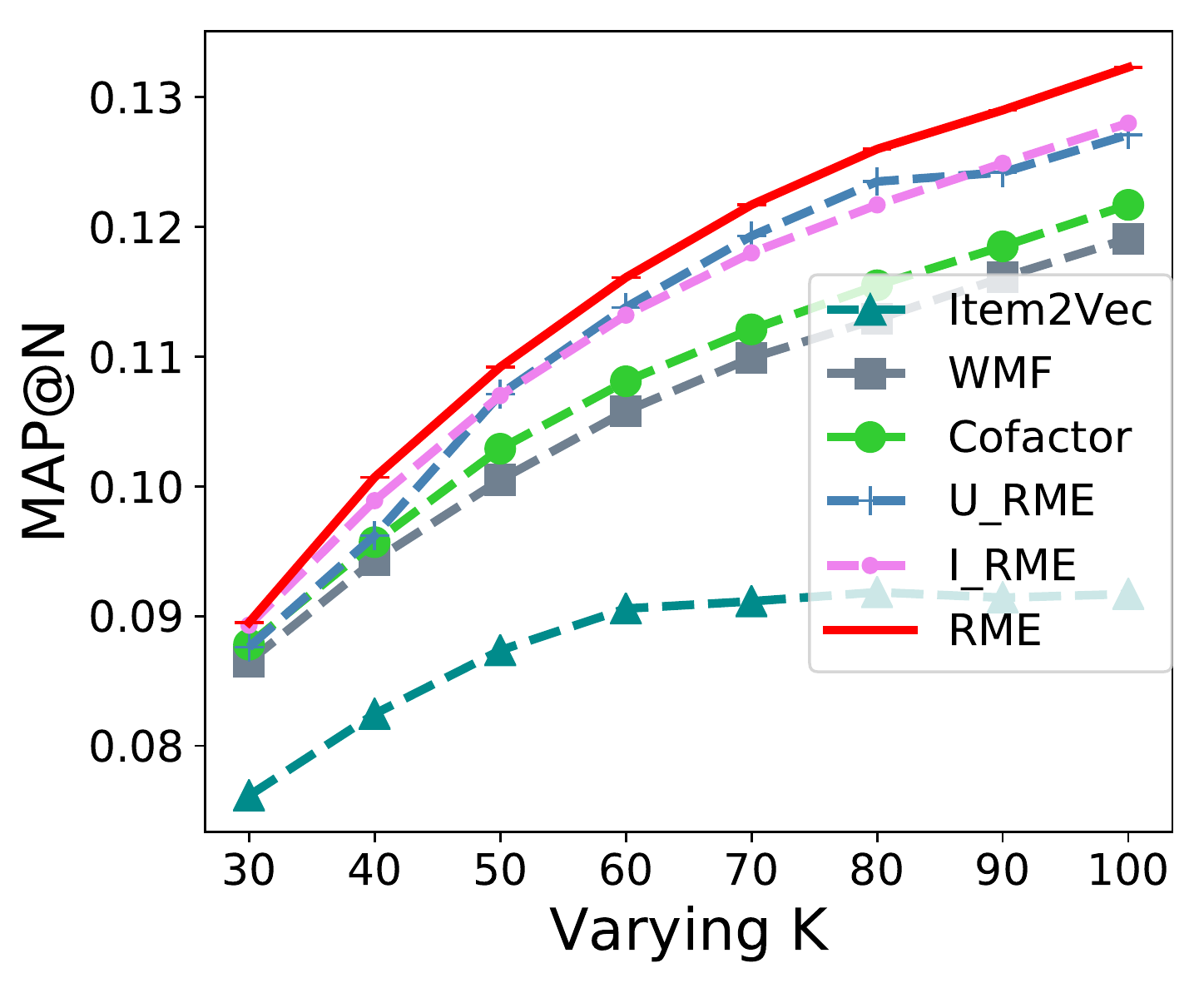}
				}\vspace{-5pt}
				\caption{Performance of models when varying the latent dimension size $k$ with fixing the value of $\lambda$.}
				\label{fig:ps-vary-k}
				\vspace{-15pt}
			\end{figure}
			
			\noindent\textbf{RQ1: Performance of the baselines and RME.}
			Table \ref{table:PerformanceComparison} presents recommendation results of RME and compared models at Recall@5, NDCG@20, and MAP@10. First, we compared RME with the baselines. We observed that RME outperformed all baselines in the three datasets, improving the Recall by 6.3\%, NDCG by 5.1\%, and MAP by 8.3\% on average over the best baseline (\emph{p-value} $<$ 0.001). Second, we compared two variants of RME model with the baselines. We see that both U\_RME and I\_RME performed better than the baselines. Adding user embeddings improved the Recall by 3.0$\sim$3.5\%, NDCG by 1.4$\sim$2.2\%, and MAP by 4.2$\sim$5.8\% (\emph{p-value} $<$ 0.001), while adding disliked item embeddings improved the Recall by 3.1$\sim$3.8\%, NDCG by 2.2$\sim$3.0\%, and MAP by 4.9$\sim$6.0\%. Third, we compare RME with its two variants. RME also achieved the best result, improving Recall by 2.6$\sim$3.0\%, NDCG by 2.0$\sim$2.5\%, and MAP by 2.6$\sim$2.8\% (\emph{p-value} $<$ 0.05). We further evaluated NDCG@N of our model when varying top $N$ in range \{5, 10, 20, 50, 100\}. Figure \ref{fig:vary-N} shows our result (we excluded Item-KNN in the figure and following figures since it performed extremely worst). Our model still performed the best. On average, it improved NDCG@N by 6.2\% comparing to the baselines, and by 3.3\% comparing to its variants. These experimental results show that both co-disliked item embedding and user embedding positively contributed to RME, and also confirm our observations addressed in Section~\ref{sec:introduction} are correct.
			
			\begin{figure}
				\centering
				\subfigure[Recall@5, NDCG@5, MAP@5 on MovieLens-10M. Fix $k=40$, and vary $\lambda$.] 
				{
					\label{fig:ml10m-vary-lambda}
					\includegraphics[width=0.155\textwidth]{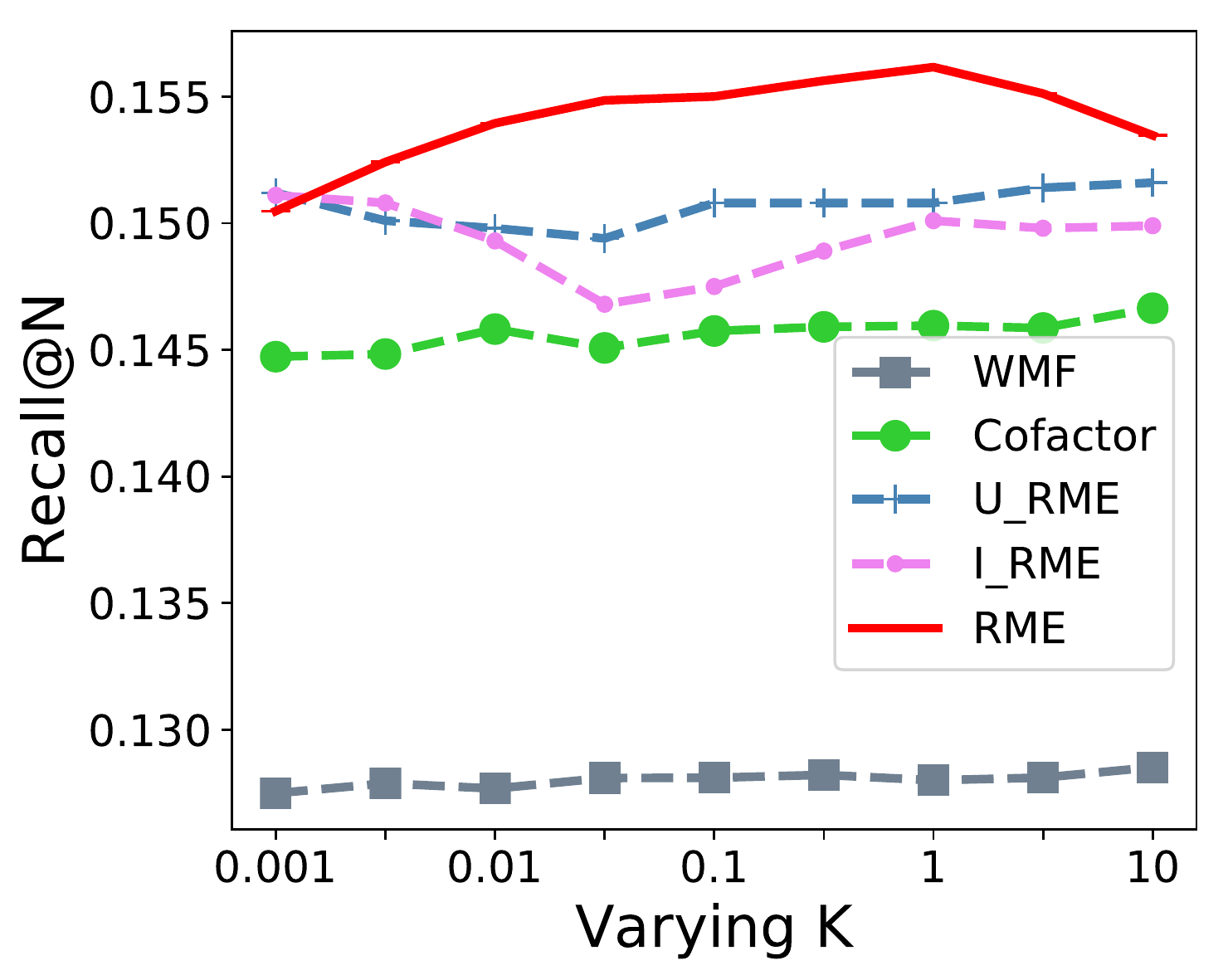}
					\includegraphics[width=0.155\textwidth]{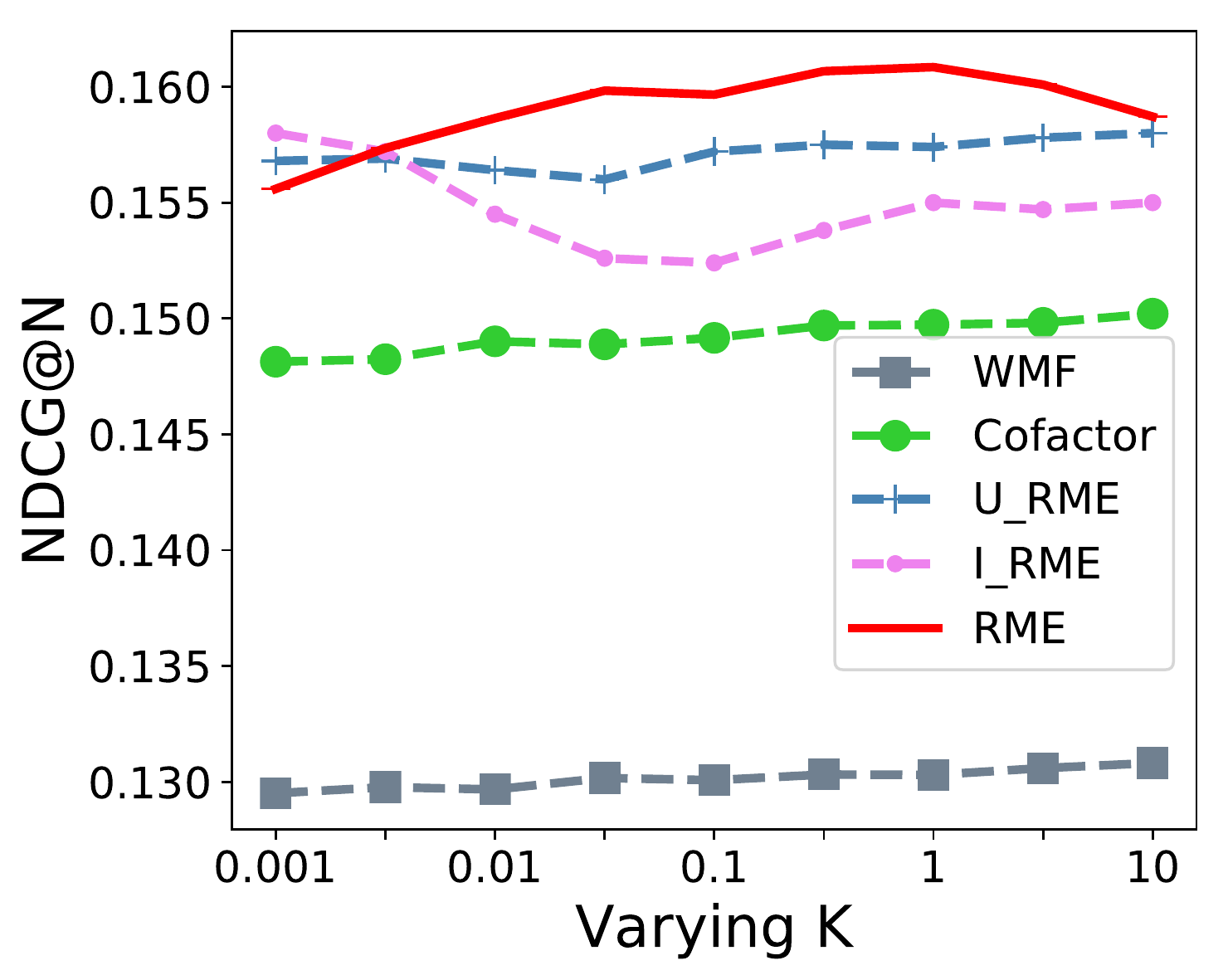}
					\includegraphics[width=0.155\textwidth]{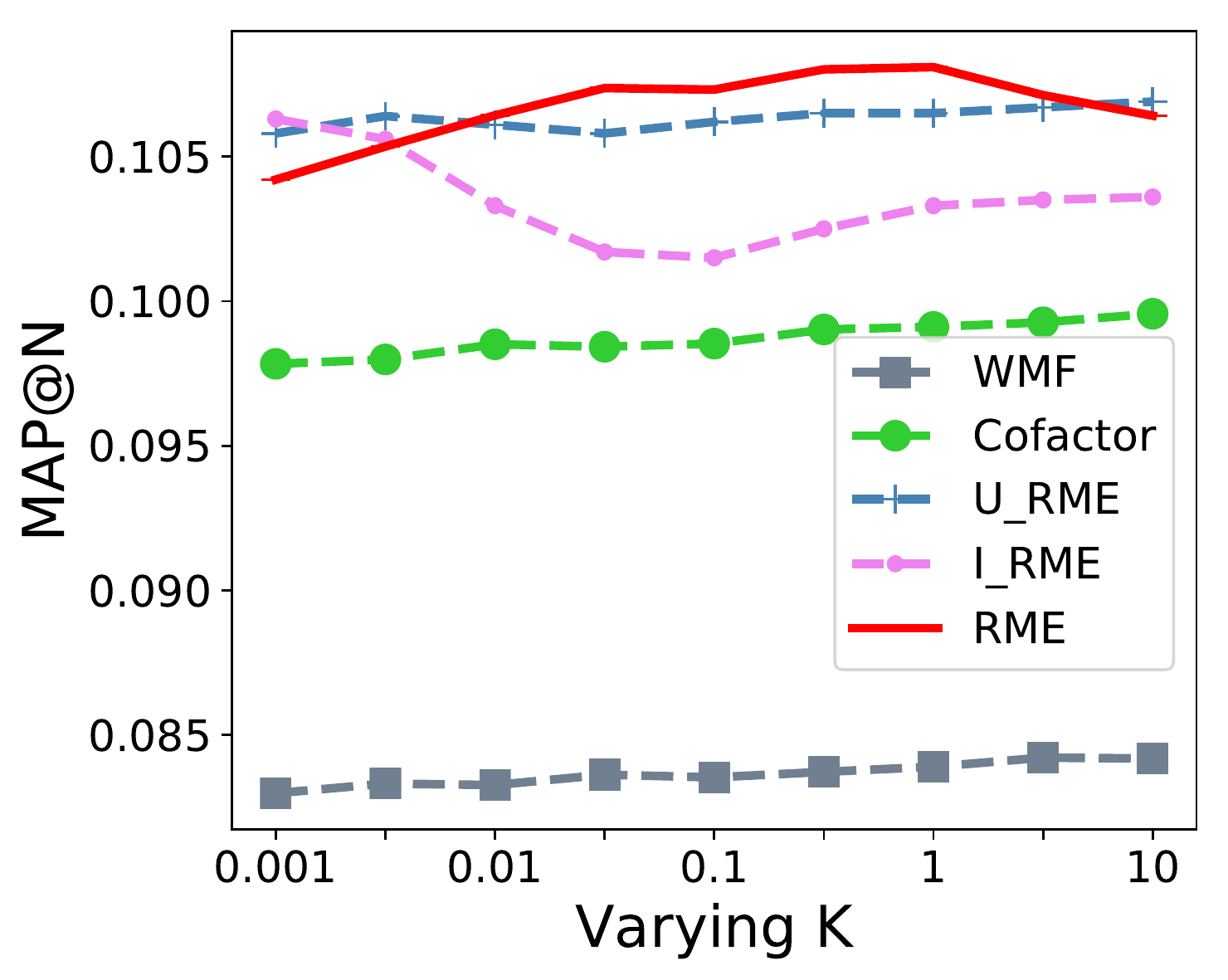}
				}\vspace{-5pt}
				\subfigure[Recall@5, NDCG@5, MAP@5 on MovieLens-20M. Fix $k=40$, and vary $\lambda$.] 
				{
					\label{fig:ml20m-vary-lambda}
					\includegraphics[width=0.155\textwidth]{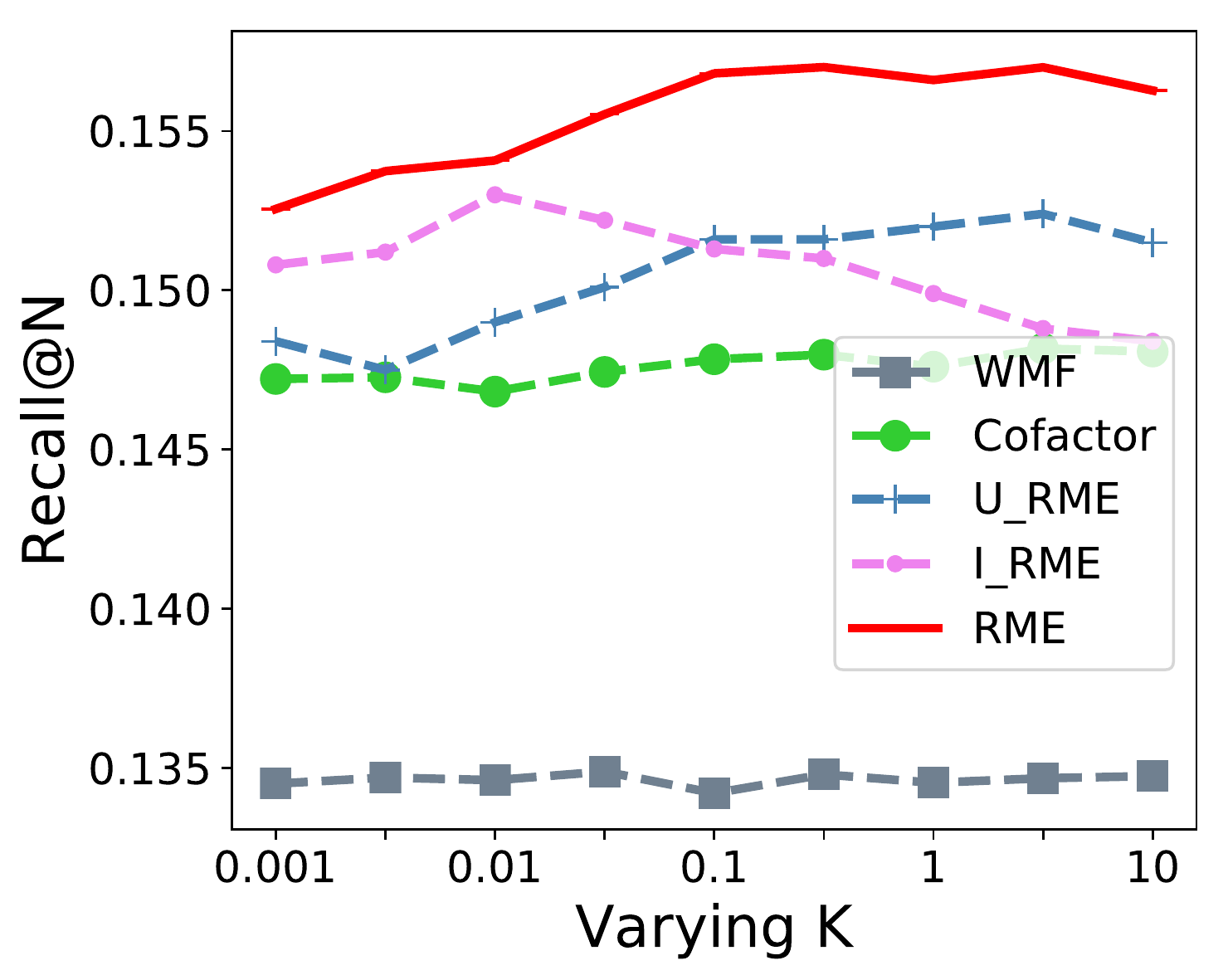}
					\includegraphics[width=0.155\textwidth]{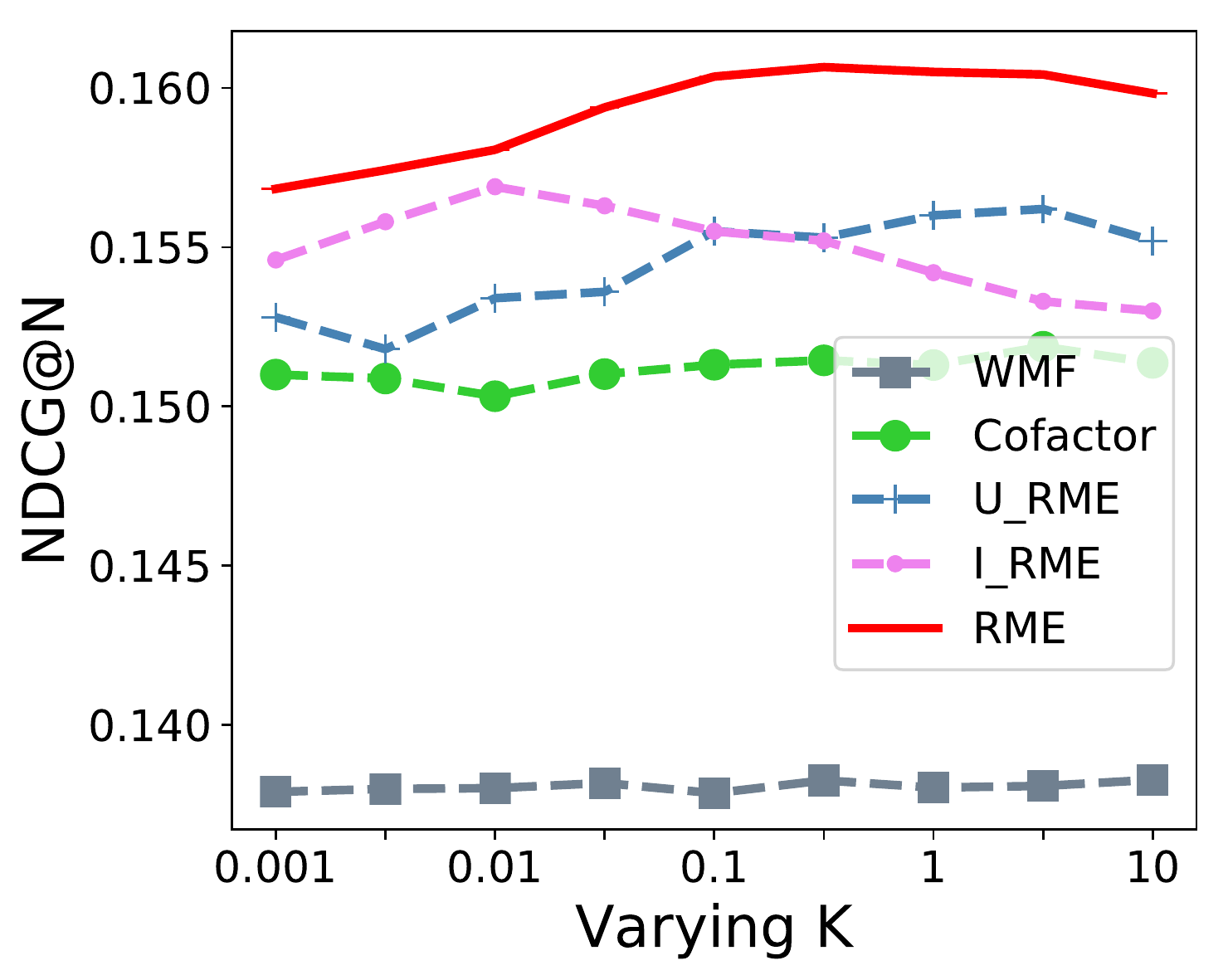}
					\includegraphics[width=0.155\textwidth]{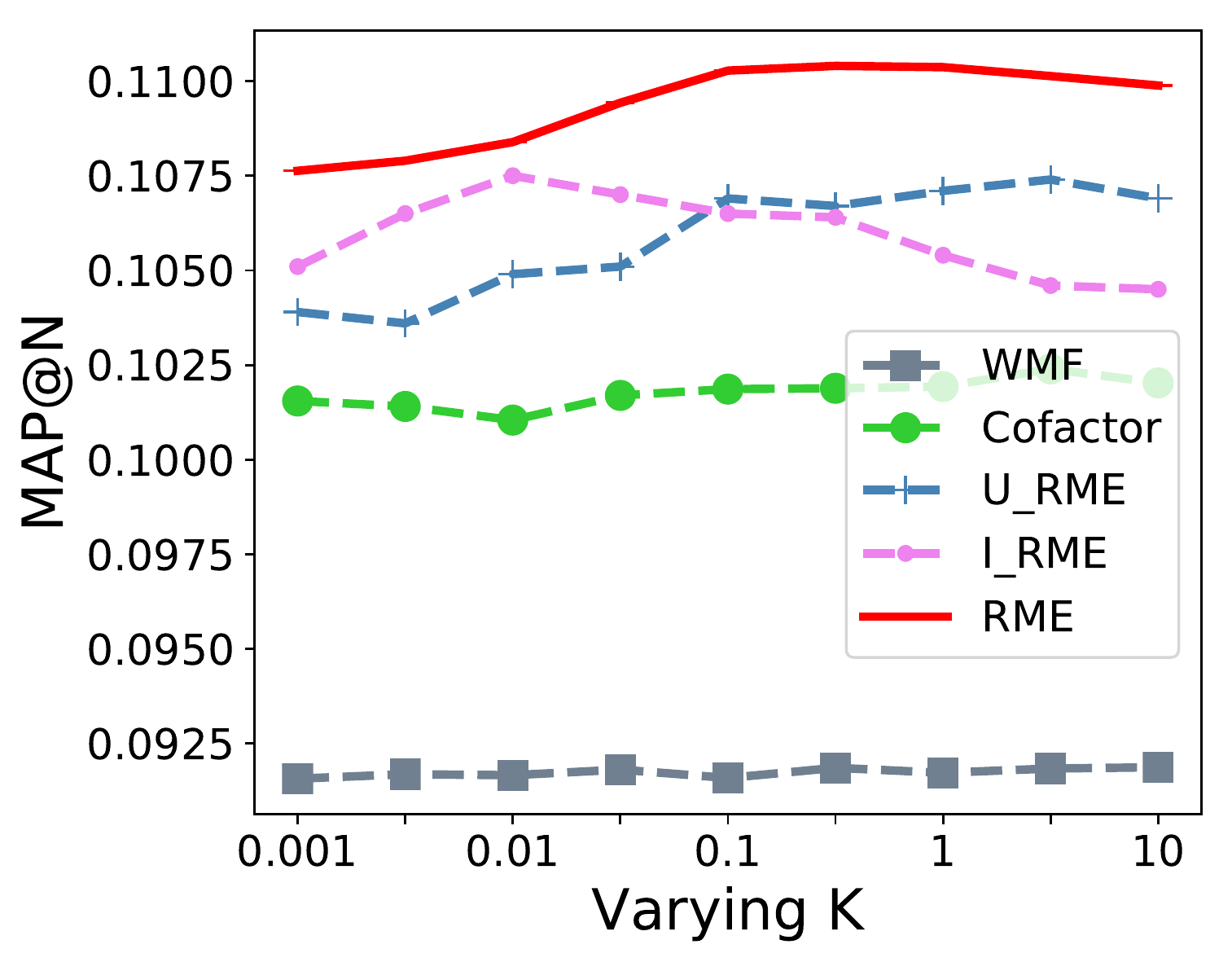}		
				}\vspace{-5pt}
				\subfigure[Recall@5, NDCG@5, MAP@5 on TasteProfile. Fix $k=100$, $\tau=0.2$, and vary $\lambda$.] 
				{
					\label{fig:tp-vary-lambda}
					\includegraphics[width=0.155\textwidth]{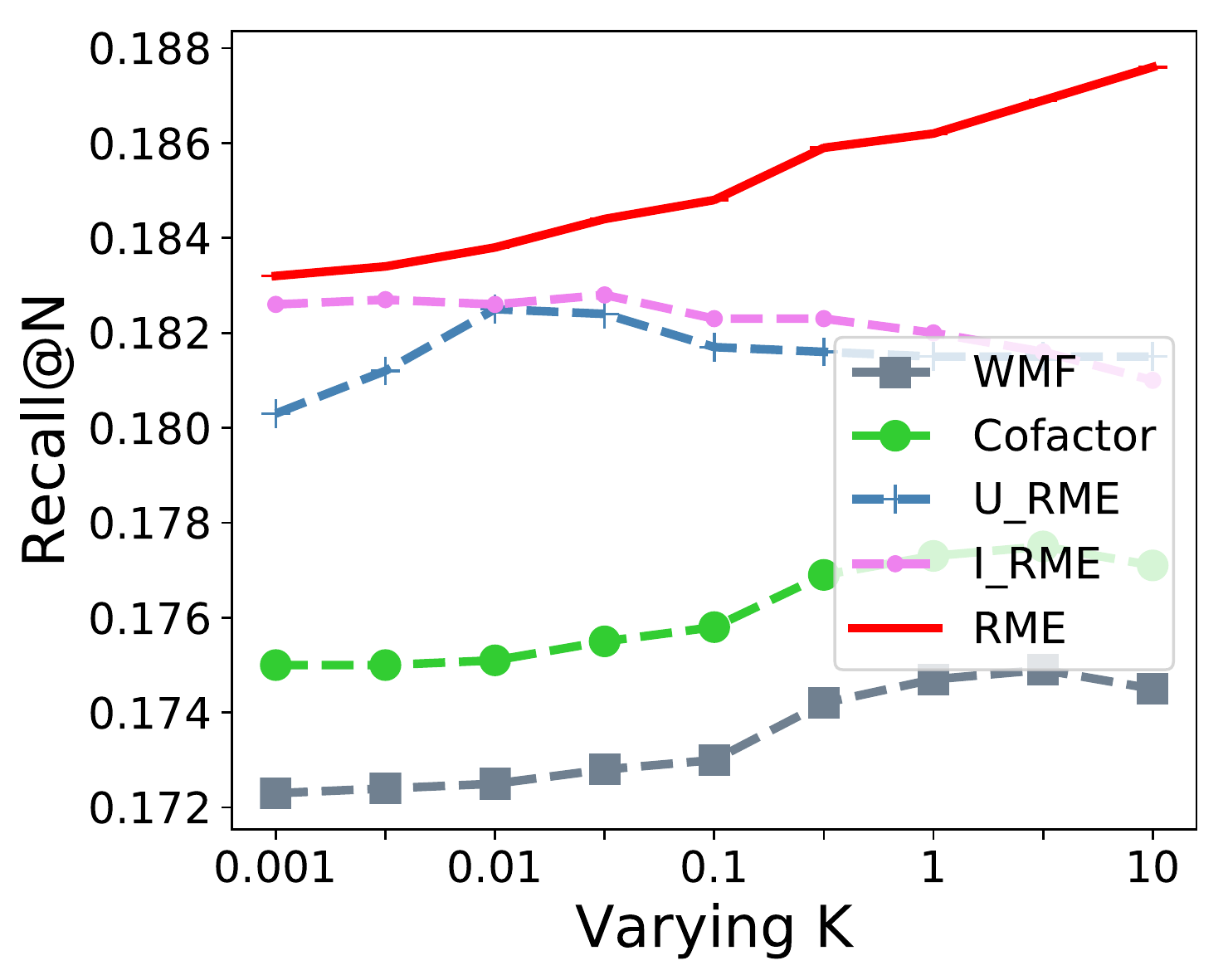}
					\includegraphics[width=0.155\textwidth]{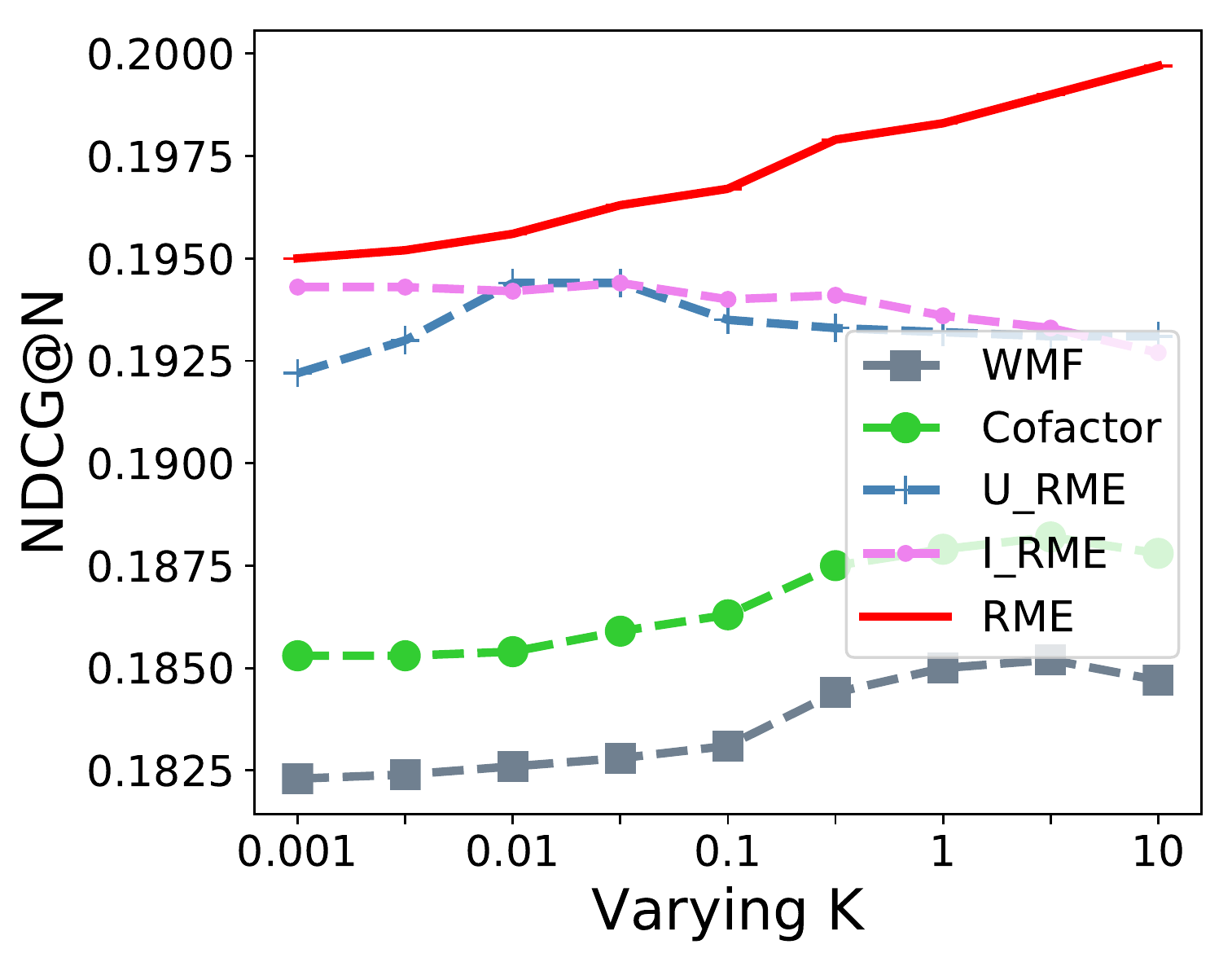}
					\includegraphics[width=0.155\textwidth]{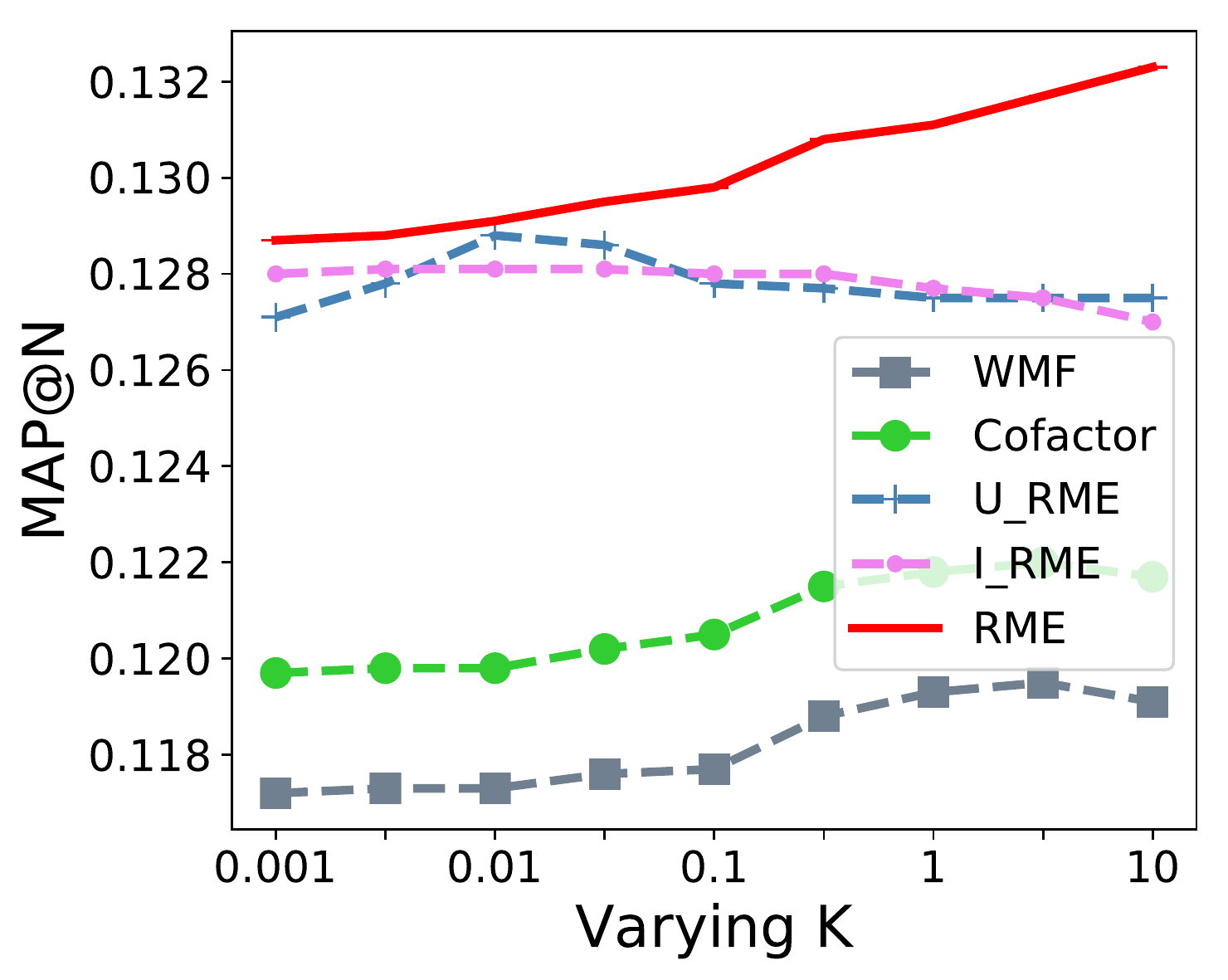}
				}\vspace{-5pt}
				\caption{Performance of models when varying $\lambda$ with fixing the latent dimension size $k$. Item2Vec did not contain regularization, so we excluded it.}
				\label{fig:ps-vary-lambda}
				\vspace{-10pt}
			\end{figure}
			
			The experimental results in TasteProfile in Table~\ref{table:PerformanceComparison} showed that inferring disliked items in Algorithm~\ref{alg:mcf-implicit} worked well since RME model incorporating co-disliked item embedding outperformed the baselines. To further confirm the effectiveness of the algorithm, we also applied it to MovieLens-10M and MovieLens-20M datasets after removing the explicit disliking information, pretending them as implicit feedback datasets. In the datasets without disliking information, RME under Algorithm~\ref{alg:mcf-implicit} still outperformed the best baseline with 4.2\%, 4.6\% and 7.2\% improvements on average in Recall, NDCG and MAP, respectively (\emph{p-value} < 0.001). Its performance was slightly lower than the original RME (based on explicit disliking information) at 0.3\%, 0.7\% and 1.3\% on average in Recall, NDCG, and MAP, respectively. The experimental results confirmed the effectiveness of Algorithm~\ref{alg:mcf-implicit}. We note that Algorithm~\ref{alg:mcf-implicit} got converged in up to 4 iterations for all three datasets by the early stopping condition. Due to the space limitation, we do not include figures which show the loss over iterations.

			\vspace{0.05in}
			\noindent\textbf{RQ2:Parameter sensitivity analysis:} We analyze the effects of the parameters in RME model in order to answer the following research questions: (\emph{RQ2-1}:) How does RME work when varying the latent dimension size $k$?; (\emph{RQ2-2}:) How does RME model change with varying $\lambda$?; (\emph{RQ2-3}:) How sensitive is the RME model on an implicit feedback dataset (e.g. TasteProfile) when varying negative sample drawing ratio $\tau$?; and (\emph{RQ2-4}:) Can RME achieve better performance with a dynamic setting of regularization hyper-parameters?
			
			Regarding \emph{RQ2-1}, Figure \ref{fig:ps-vary-k} shows the sensitivity of all compared models when fixing $\lambda$ and varying the latent dimension size $k$ in \{30, 40, 50, 60, 70, 80, 90, 100\}. It is clearly observed that our model outperforms the baselines in all datasets. In MovieLens-10M and MovieLens-20M datasets, all six models downgrade the performance when the latent dimension size $k$ is over 60. In the TasteProfile dataset, when increasing $k$, although all models gain a higher performance, our model tends to achieve much higher performance.
			
			In a \emph{RQ2-2} experiment, we exclude Item2Vec because this model does not contain the regularization term. We fix $k=40$ in MovieLens-10M and MovieLens-20M. In TasteProfile dataset, we fix $k$=100, $\tau$=0.2. We vary lambda in range \{0.001, 0.005, 0.01, 0.05, 0.1, 0.5, 1, 5, 10\}. Then, we report the average results of Recall@5, NDCG@5, and MAP@5. As shown in Figure \ref{fig:ps-vary-lambda}, the performance of our model is better than the baselines. In MovieLens-10M and MovieLens-20M dataset, RME increases its performance when increasing $\lambda$ up to 1, then its performance goes down when $\lambda$ is increasing more. In TasteProfile, RME tends to gain a higher performance and more outperformed the baselines when $\lambda$ is increasing.

\begin{figure}
				{
					\includegraphics[width=0.155\textwidth]{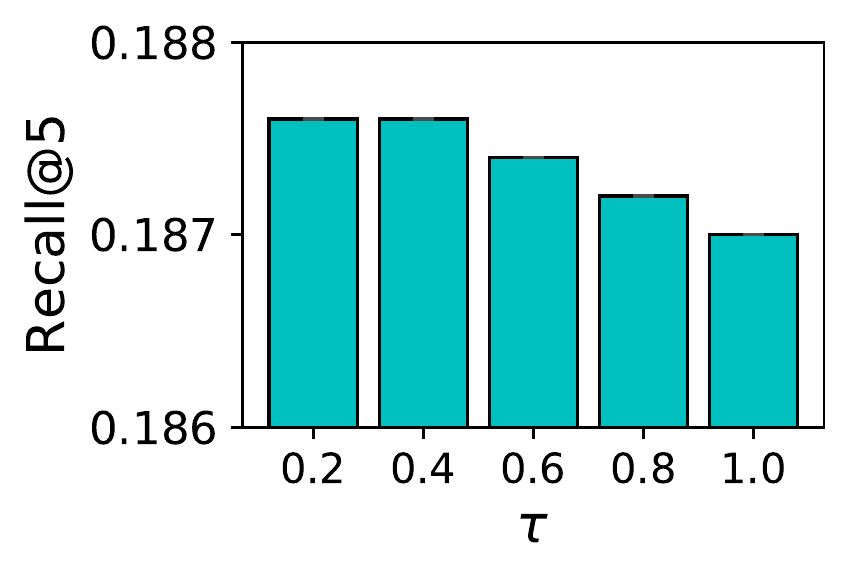}
					\includegraphics[width=0.155\textwidth]{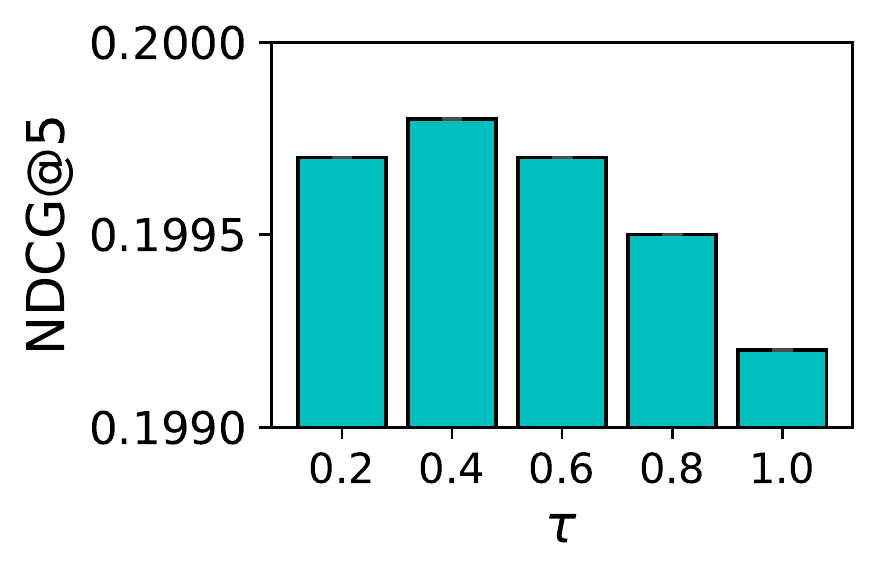}
					\includegraphics[width=0.155\textwidth]{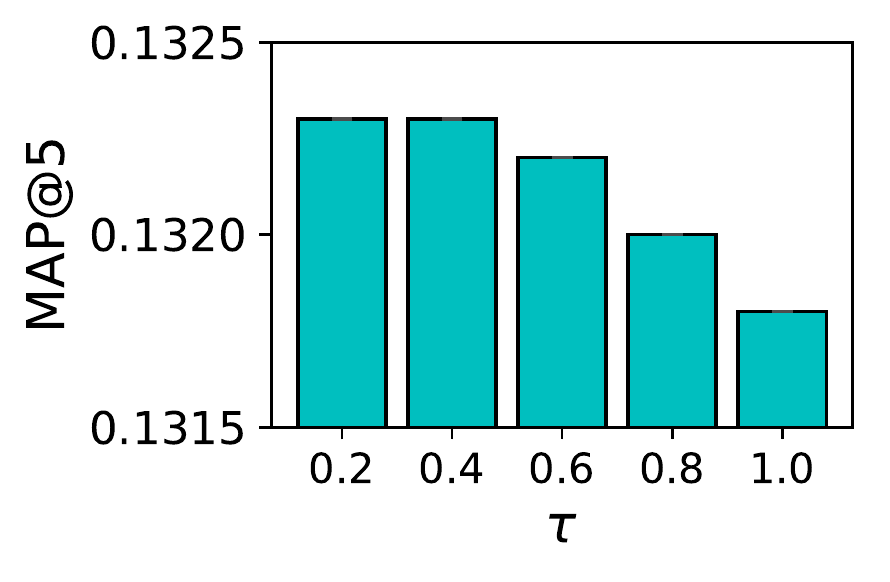}
				}
				\vspace{-15pt}
				\caption{Performance of RME in TasteProfile when varying negative sample drawing ratio $\tau$ with fixing $k=100$ and $\lambda=10$.}
				\vspace{-15pt}
				\label{fig:ps-tp-neg-ratio}
			\end{figure}

			To understand the sensitivity of our model when varying negative sample drawing ratio $\tau$ in the implicit feedback dataset -- TasteProfile (RQ2-3), we vary $\tau$ in \{0.2, 0.4, 0.6, 0.8, 1.0\}, and fix $k=100$ and $\lambda=10$. Figure \ref{fig:ps-tp-neg-ratio} shows that when $\tau$ increases, our model degrades with a small amount (e.g. around -0.3\% in Recall@5 and NDCG@5, and -0.4\% in MAP@5). In NDCG@5, our model gains the best result when $\tau=0.4$. We note that our worst case (when $\tau=1.0$) is still better than the best baseline presented in Table \ref{table:PerformanceComparison}. This shows that the sensitivity of our model with regard to the negative sample drawing ratio $\tau$ is small/limited.
			
\begin{figure}
				\subfigure[MovieLens-10M]
				{
					\includegraphics[width=0.21\textwidth]{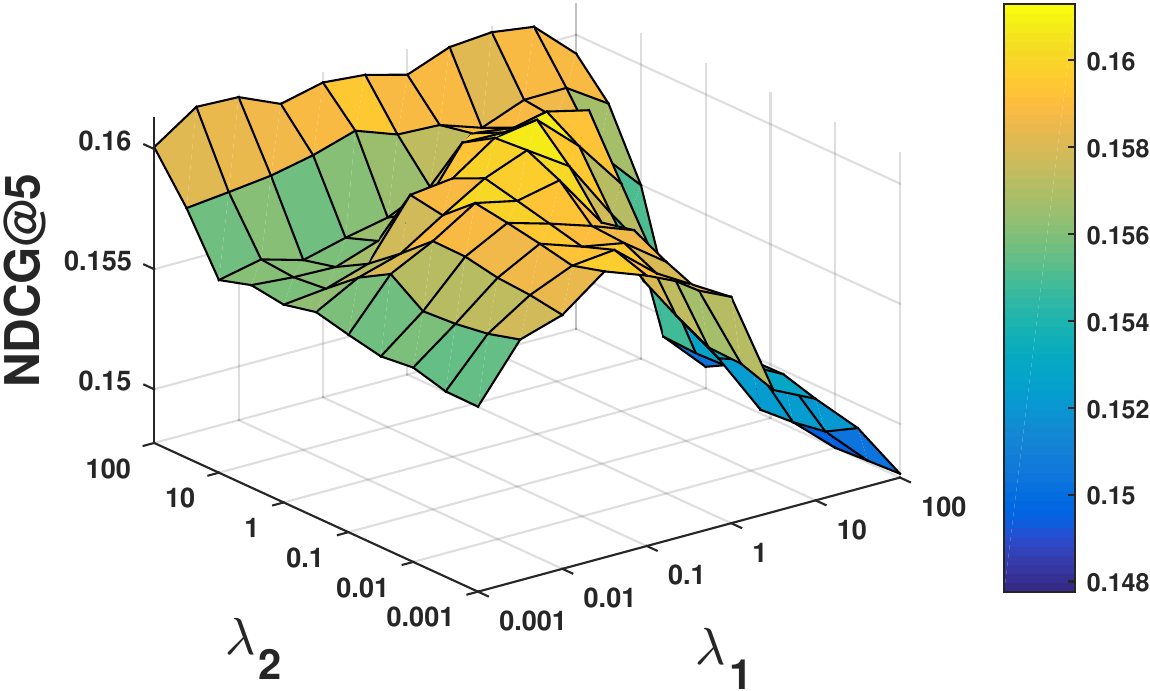}
				}
				\subfigure[MovieLens-20M]
				{
					\includegraphics[width=0.21\textwidth]{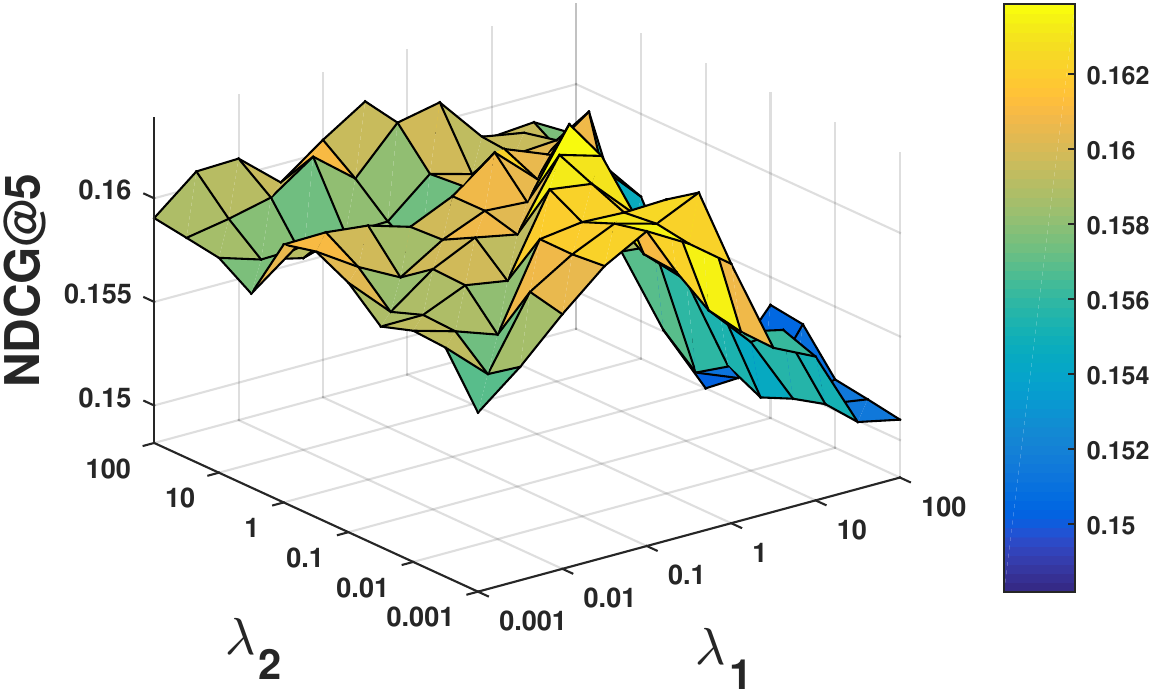}
				}\vspace{-5pt}
				\subfigure[TasteProfile]
				{
					\includegraphics[width=0.21\textwidth]{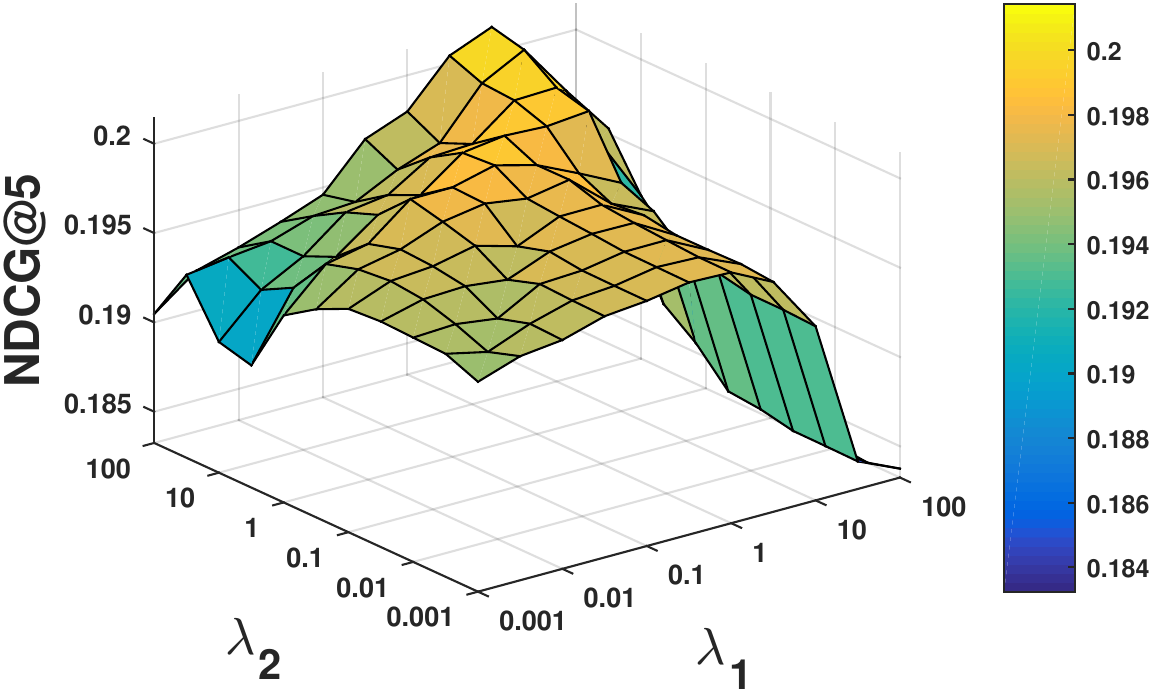}
				}
				\vspace{-5pt}
				\caption{Performance of RME under a dynamic setting of regularization hyper-parameters. Set $\lambda_{\alpha}$ = $\lambda_{\beta}$ = $\lambda_1$, and  $\lambda_{\gamma^{(+)}}$ =  $\lambda_{\gamma^{(-)}}$ = $\lambda_{\theta}$ = $\lambda_2$.}
				\label{fig:3d-dynamic-settings}
				\vspace{-10pt}
			\end{figure}

			In our previous experiments, we used a static setting of regularization hyper-parameters by setting $\lambda_\alpha = \lambda_\beta = \lambda_{\gamma} = \lambda_{\delta} = \lambda_\theta = \lambda$. To explore if a dynamic setting of those regularization hyper-parameters could lead to better results for RME model (RQ2-4), we set $\lambda_\alpha = \lambda_\beta = \lambda_1$, $\lambda_{\gamma} = \lambda_{\delta} = \lambda_\theta = \lambda_2$. Then we both vary $\lambda_1$ and $\lambda_2$ in \{100, 50, 10, 5, 1, 0.5, 0.1, 0.05, 0.01, 0.005, 0.001\} while fixing the latent dimension size $k$. Next, we report the NDCG@5 for all 3 datasets. As shown in Figure \ref{fig:3d-dynamic-settings}, our model even get a higher performance with the dynamic setting. For example, it gains NDCG@5 = 0.1613 when $\lambda_1 = 100$ and $\lambda_2 = 0.005$ in MovieLens-10M dataset. Similarly, NDCG@5 = 0.1639 when $\lambda_1 = 0.5$, $\lambda_2 = 1$ in MovieLens-20M dataset. NDCG@5 = 0.2014 when $\lambda_1 = 100$, $\lambda_2 = 10$ in TasteProfile dataset. The dynamic setting produced 0.3$\sim$2\% higher results than the static setting presented in Table \ref{table:PerformanceComparison}.

			So far, we compared the performance of our model and the baselines while varying values of hyper-parameters. We showed that our model outperformed the baselines in all cases, indicating that our model was less sensitive with regard to the hyper-parameters. We also showed that our model produced better results under the dynamic setting.
			
			\begin{figure*}
				\centering
				\subfigure[Dataset: MovieLens-10M. Performance of RME over other models in all three user groups are significant (\emph{p-value} < 0.05).]
				{
					\label{fig:ml10m-activeness}
					\includegraphics[width=0.22\textwidth]{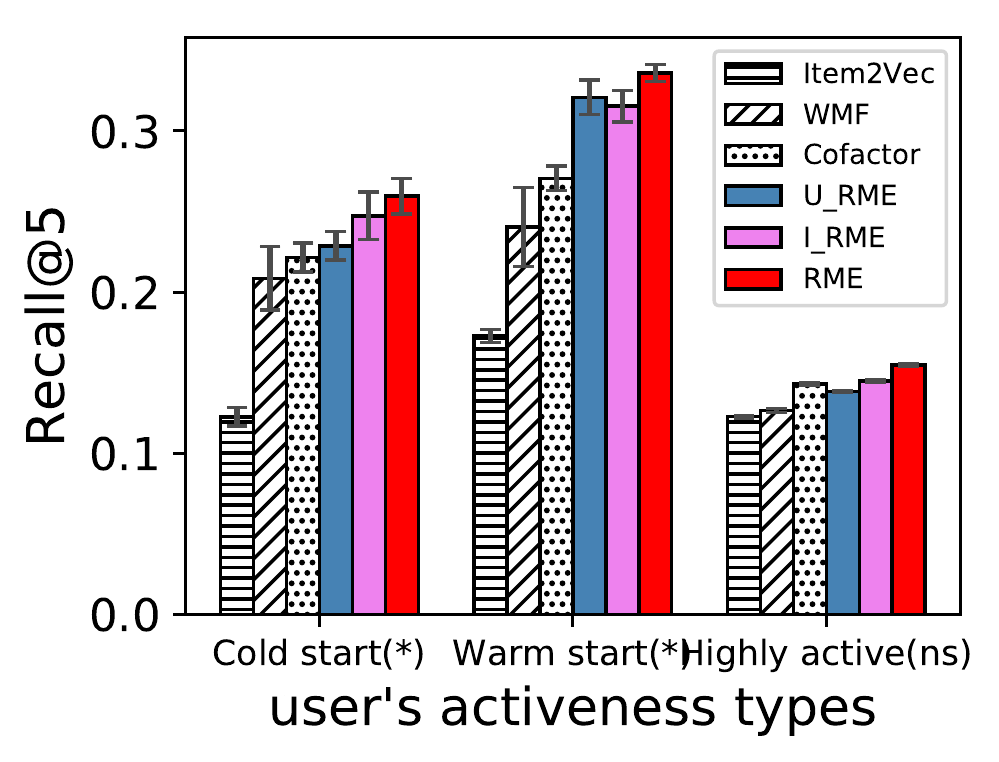}
					\includegraphics[width=0.22\textwidth]{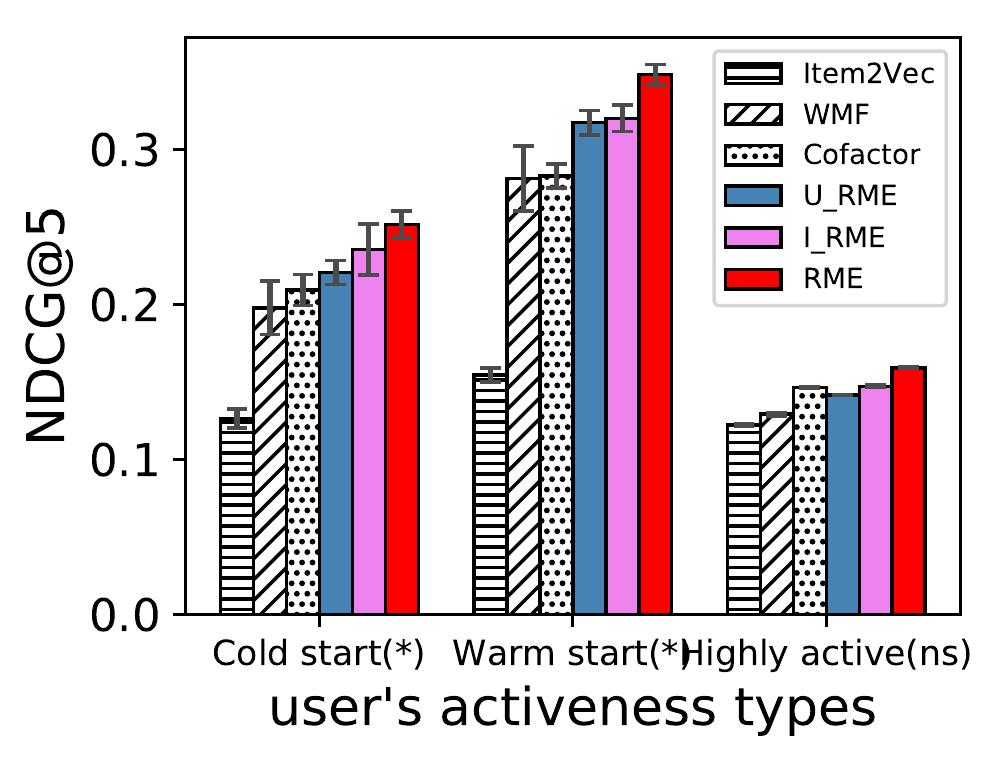}
					\includegraphics[width=0.22\textwidth]{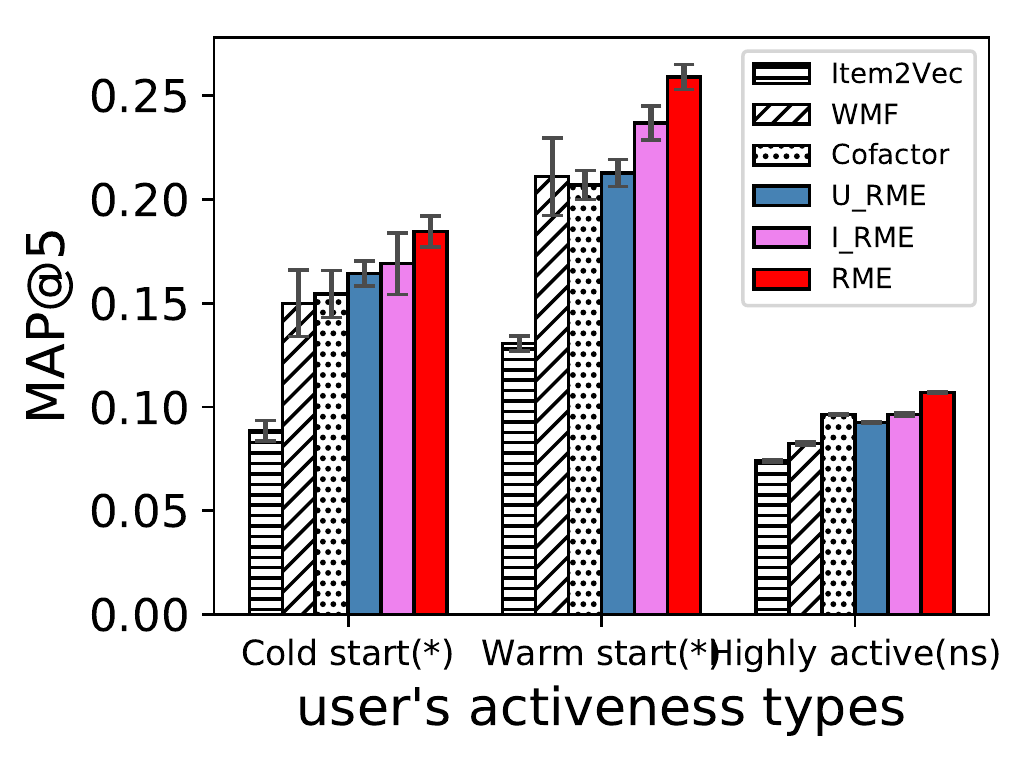}
				}
				\vspace{-10pt}
				
				\subfigure[Dataset: MovieLens-20M. Performance of RME over other models in \emph{cold-start} and \emph{highly-active} user groups are significant (\emph{p-value} < 0.05).]
				{
					\label{fig:ml20m-activeness}
					\includegraphics[width=0.22\textwidth]{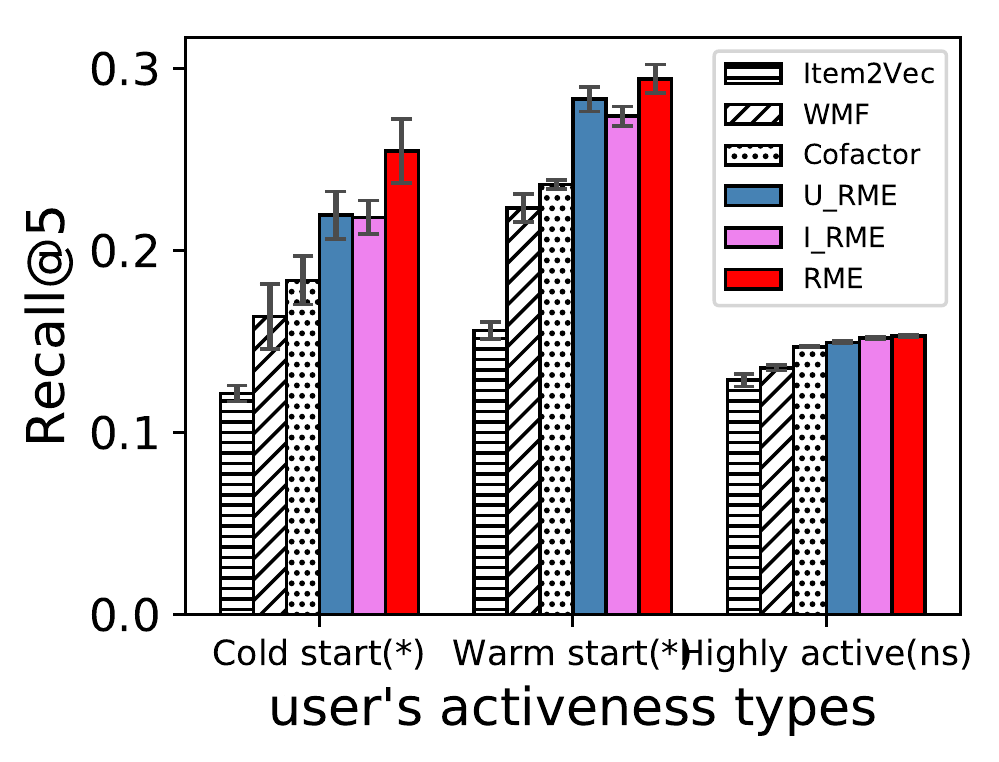}
					\includegraphics[width=0.22\textwidth]{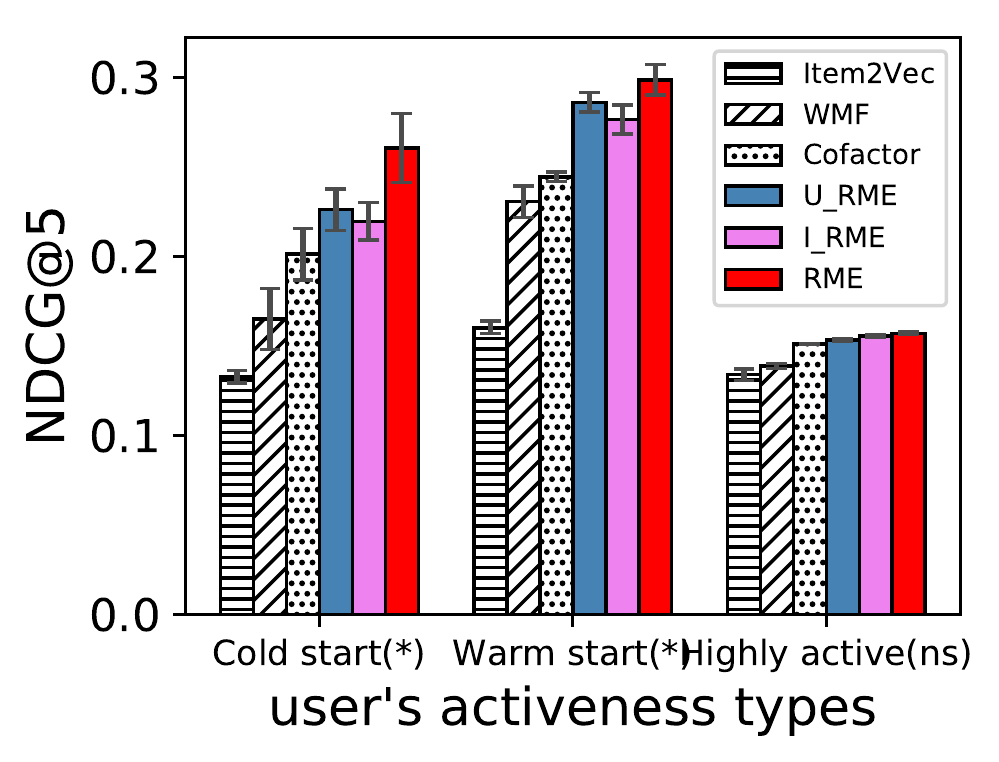}
					\includegraphics[width=0.22\textwidth]{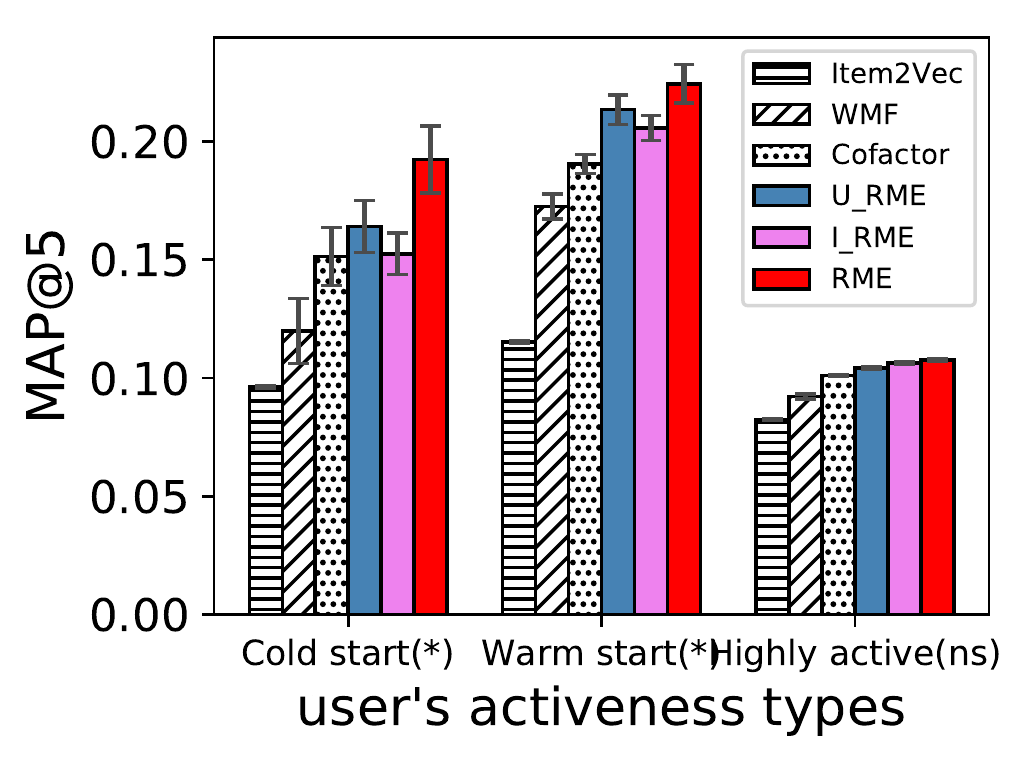}
				}
				\vspace{-10pt}		
				
				\subfigure[Dataset: TasteProfile. Performance of RME over other models in \emph{highly-active} user group is significant (\emph{p-value} < 0.05).]
				{
					\label{fig:tp-activeness}
					\includegraphics[width=0.22\textwidth]{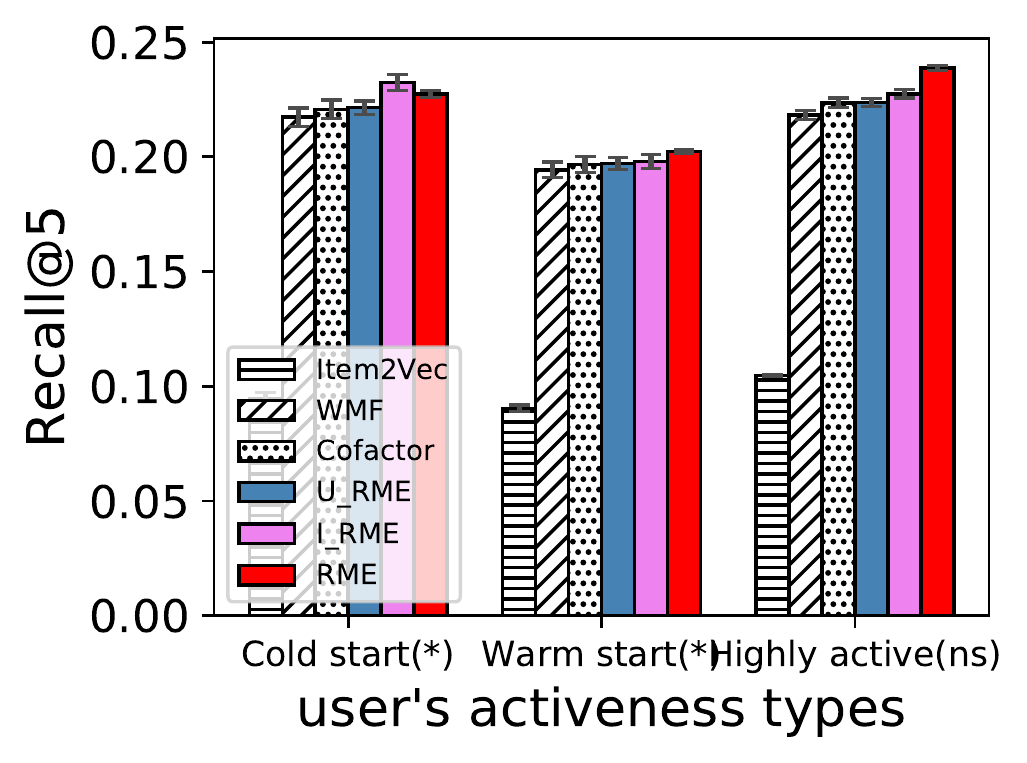}
					\includegraphics[width=0.22\textwidth]{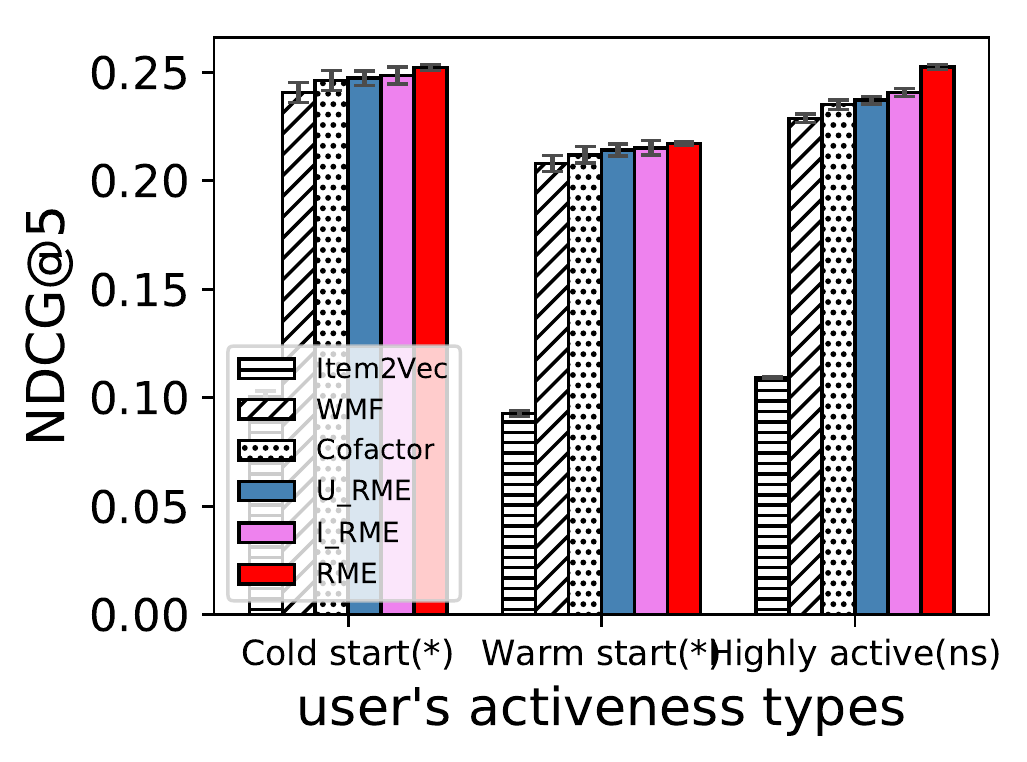}
					\includegraphics[width=0.22\textwidth]{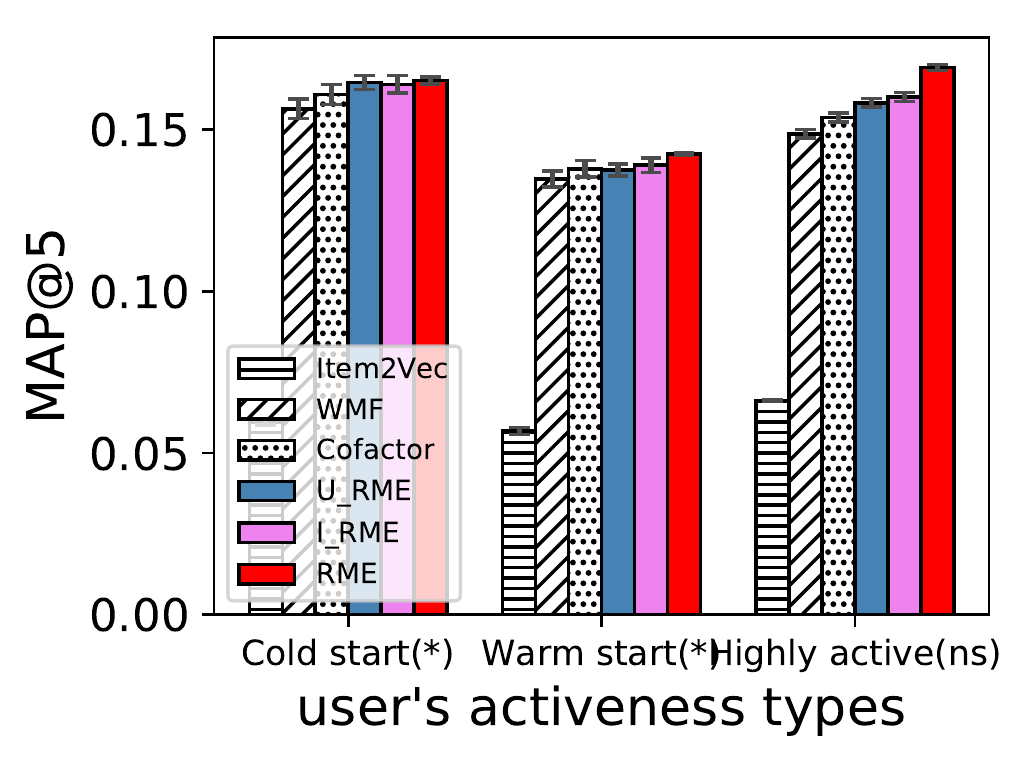}
				}
				\vspace{-10pt}
				\caption{Performance of models for each of the three user groups. Non-directional two-sample t-test was performed. * indicates significant (\emph{p-value} $<$ 0.05), and \emph{ns} indicates not significant. The error bars are the average of standard errors in the 5 folds.} 
				\label{fig:user-activeness}
				\vspace{-10pt}
			\end{figure*}

			\vspace{0.05in}
			\noindent\textbf{RQ3: Performance of models for different types of users.} We sorted users by the ascending order of their activity level in terms of the number of liked items. Then we categorized them into three groups: (1) \emph{cold-start users} who were in the first 20\% of the sorted user list (i.e., their activity level is the lowest); (2) \emph{warm-start users} who were in between 20\% and 80\% of the sorted user list; (3) \emph{highly active users} who were in the last 20\% of the sorted user list (i.e., the most active users). Then, we measured the performance of all the compared models for each of the user groups.
			
			Figure \ref{fig:user-activeness} shows the performance of all the compared models in MovieLens-10M, MovieLens-20M and TasteProfile datasets. In MovieLens-10M (Figure \ref{fig:ml10m-activeness}), our model significantly outperformed the baselines and the two variants in all three user groups, improving Recall@5 by 4.7$\sim$6.7\%, NDCG@5 by 6.8$\sim$8.8\%, and MAP@5 by 9.2$\sim$11.0\% over the best compared method. In MovieLens-20M dataset (Figure \ref{fig:ml20m-activeness}), our model significantly outperformed the baselines and its variants in 2 groups: \emph{cold-start users} and \emph{warm-start users}. It improved Recall@5 by 16.1\%, 4.0\%, 0.8\%, NDCG@5 by 15.3\%, 4.4\%, 0.9\%, MAP@5 by 17.3\%, 5.1\%, 1.1\% in \emph{cold-start users}, \emph{warm-start users} and \emph{highly-active users}, respectively. Specially, in both MovieLens-10M and MovieLens-20M datasets, our model on average much improved the baselines in \emph{cold-start users} with Recall@5, NDCG@5 and MAP@5 by 27.9\%, 24.8\% and 23.3\%, respectively. It shows the benefit of incorporating disliked item embeddings and user embeddings. In TasteProfile dataset (Figure \ref{fig:tp-activeness}), our model significantly improved baselines in \emph{highly-active users} group, improving Recall@5 by 6.8\%, NDCG@5 by 7.4\%, and MAP@5 by 10.0\% comparing to the best state-of-the-art method, while improving Recall@5 by 5.0\%, NDCG@5 by 4.9\%, and MAP@5 by 5.7\% comparing to its best variant. However, in \emph{cold-start users} and \emph{warm-start users} group, RME got an equal performance comparing with the baselines (i.e., the difference between our model and other methods are not significant). 

			\vspace{0.05in}
			\noindent\textbf{RQ4: Joint learning vs separate learning.} What if we conduct learning separately for each part of our model? Will the separate learning model perform better than our joint learning model? To answer the questions, we built a separate learning model as follows: first, we learned latent representations of items by jointly decomposing two SPPMI matrices $X^{(+)}$ and $X^{(-)}$ of liked item-item co-occurrences and disliked item-item co-occurrences, respectively. Then, we learned user's latent representations by minimizing the objective function in Equation (\ref{func:obj-model5}), where the latent representations of items and item contexts were already learned and fixed. Next, we compared our joint learning model (i.e., RME) with the separate learning model in MovieLens-10M, MovieLens-20M, and TasteProfile datasets. Our experimental results show that our joint learning model outperformed the separate learning model by significantly improving Recall@5, NDCG@5 and MAP@5 at least 12.1\%, 13.5\% and 17.1\%, respectively (p-value $<$ 0.001).

			\section{Related Work}
			\label{sec:relatedworks}

			\noindent\textbf{Latent factor models (LFM):} Some of the first works in recommendation focused on explicit feedback datasets (name some: \cite{salakhutdinov2007restricted,sarwar2001item,koren2008factorization}). Our proposed method worked well for both explicit and implicit feedback settings with almost equal performances.
			
			In implicit feedback datasets, which have been trending recently due to the difficulty of collecting users' explicit feedback, properly treating/modeling missing data is a difficult problem \cite{bayer2017generic,liang2016modeling,pilaszy2010fast}. Even though missing values are a mixture of negative feedback and unknown feedback, many works treated all missing data as negative instances \cite{devooght2015dynamic,hu2008collaborative,pilaszy2010fast,volkovs2015effective}, or sampled missing data as negative instances with uniform weights \cite{rendle2009bpr}. This is suboptimal because treating negative instances and missing data differently can further improve recommenders' performance \cite{he2016fast}. \cite{pan2008one} proposed a bagging of ALS learners \cite{hu2008collaborative} to sample negative instances. He et al. \cite{he2016fast} assumed that unobserved popular items have a higher chance of being negative instances. In our work, we attempted to non-uniformly sample negative instances in implicit feedback datasets and treated them as additional information to enhance our model performance. Specifically, we (i) designed an EM-like algorithm with a softmax function to draw personalized negative instances for each user; (ii) employed a word embedding technique to exploit the co-occurrence patterns among disliked items, further enriching their latent representations.
			
			
			\noindent\textbf{LFM with auxiliary information:} In latent factor models, additional sources of information were incorporated to improve collaborative filter-based recommender systems (e.g., user reviews, item categories, and article information \cite{almahairi2015learning,guardia2015latent,mcauley2013hidden,wang2011collaborative}). However, we only used an user-item-preference matrix without requiring additional side information. Adding the side information into our model would potentially further improve its performance. But, it is not a scope of our work in this paper.
			
			\noindent\textbf{LFM with item embeddings:} \cite{wang2017incorporating} incorporated message embedding for retweet prediction. Cao et al. \cite{cao2017embedding} co-factorized the user-item interaction matrix, user-list interaction matrix, and item-list co-occurrences to recommend songs and lists of songs for users. \cite{liang2016factorization} learned liked item embeddings with an equivalent matrix factorization method of skip-gram negative sampling (SGNS), and performed joint learning with matrix factorization. \cite{barkan2016item2vec} exploited item embeddings using the SGNS method for item collaborative filtering. So far, the closest techniques to ours \cite{barkan2016item2vec,liang2016factorization} only considered liked item embeddings, but we proposed a joint learning model that not only considered LFM using matrix factorization with liked item embeddings, but also user embeddings and disliked item embeddings. Since integrating co-disliked item embedding is non-trivial for implicit feedback datasets, we also proposed an EM-like algorithm for extracting personalized negative instances for each user.

			\vspace{0.02in}
			\noindent\textbf{Word embeddings:} Word embedding models \cite{mikolov2013distributed,pennington2014glove} represent each word as a vector of real numbers called word embeddings. In \cite{levy2014neural}, the authors proposed an implicit matrix factorization that was equivalent to word2vec \cite{mikolov2013distributed}. To extend word2vec, researchers proposed models that mapped paragraphs or documents to vectors \cite{le2014distributed,djuric2015hierarchical}. In our work, we applied word embedding techniques to learn latent representations of users and items.
			
			
			
			
			\section{Conclusion}
			\label{sec:conclusion}
			
			In this paper, we proposed to exploit different co-occurrence information: co-disliked item-item co-occurrences and user-user co-occurrences, which were extracted from the user-item interaction matrix. We proposed a joint model combining WMF, co-liked embedding, co-disliked embedding and user embedding, following the recent success of word embedding techniques. Through comprehensive experiments, we successfully demonstrated that our model outperformed all baselines, significantly improving NDCG@20 by 5.6\% in MovieLens-10M dataset, by 5.3\% in MovieLens-20M dataset, and by 4.3\% in TasteProfile dataset. We also analyzed how our model worked on different types of users in terms of their interaction activity levels. We observed that our model significantly improved NDCG@5 by 20.2\% in MovieLens-10M, by 29.4\% in MovieLens-20M for the \emph{cold-start users} group. In the future extension of our model, we are interested in selecting contexts for users/items by setting a timestamp-based window size for timestamped datasets. In addition, we are also interested in incorporating co-disliked patterns among users (i.e., co-disliked user embeddings) into our model.
			
			\section{ACKNOWLEDGMENT}
			This work was supported in part by NSF grants CNS-1755536, CNS-1422215, DGE-1663343, CNS-1742702, DGE-1820609, Google Faculty Research Award, Microsoft Azure Research Award, and Nvidia GPU grant. Any opinions, findings and conclusions or recommendations expressed in this material are the author(s) and do not necessarily reflect those of the sponsors.

			\bibliographystyle{ACM-Reference-Format}
			\bibliography{sigproc}


\begin{thebibliography}{40}


\ifx \showCODEN    \undefined \def \showCODEN     #1{\unskip}     \fi
\ifx \showDOI      \undefined \def \showDOI       #1{#1}\fi
\ifx \showISBNx    \undefined \def \showISBNx     #1{\unskip}     \fi
\ifx \showISBNxiii \undefined \def \showISBNxiii  #1{\unskip}     \fi
\ifx \showISSN     \undefined \def \showISSN      #1{\unskip}     \fi
\ifx \showLCCN     \undefined \def \showLCCN      #1{\unskip}     \fi
\ifx \shownote     \undefined \def \shownote      #1{#1}          \fi
\ifx \showarticletitle \undefined \def \showarticletitle #1{#1}   \fi
\ifx \showURL      \undefined \def \showURL       {\relax}        \fi
\providecommand\bibfield[2]{#2}
\providecommand\bibinfo[2]{#2}
\providecommand\natexlab[1]{#1}
\providecommand\showeprint[2][]{arXiv:#2}

\bibitem[\protect\citeauthoryear{Agarwal and Chen}{Agarwal and Chen}{2009}]%
        {agarwal2009regression}
\bibfield{author}{\bibinfo{person}{Deepak Agarwal} {and}
  \bibinfo{person}{Bee-Chung Chen}.} \bibinfo{year}{2009}\natexlab{}.
\newblock \showarticletitle{Regression-based latent factor models}. In
  \bibinfo{booktitle}{{\em SIGKDD}}. \bibinfo{pages}{19--28}.
\newblock


\bibitem[\protect\citeauthoryear{Almahairi, Kastner, Cho, and
  Courville}{Almahairi et~al\mbox{.}}{2015}]%
        {almahairi2015learning}
\bibfield{author}{\bibinfo{person}{Amjad Almahairi}, \bibinfo{person}{Kyle
  Kastner}, \bibinfo{person}{Kyunghyun Cho}, {and} \bibinfo{person}{Aaron
  Courville}.} \bibinfo{year}{2015}\natexlab{}.
\newblock \showarticletitle{Learning distributed representations from reviews
  for collaborative filtering}. In \bibinfo{booktitle}{{\em RecSys}}.
  \bibinfo{pages}{147--154}.
\newblock


\bibitem[\protect\citeauthoryear{Barkan and Koenigstein}{Barkan and
  Koenigstein}{2016}]%
        {barkan2016item2vec}
\bibfield{author}{\bibinfo{person}{Oren Barkan} {and} \bibinfo{person}{Noam
  Koenigstein}.} \bibinfo{year}{2016}\natexlab{}.
\newblock \showarticletitle{Item2vec: neural item embedding for collaborative
  filtering}. In \bibinfo{booktitle}{{\em MLSP Workshop}}.
  \bibinfo{pages}{1--6}.
\newblock


\bibitem[\protect\citeauthoryear{Bayer, He, Kanagal, and Rendle}{Bayer
  et~al\mbox{.}}{2017}]%
        {bayer2017generic}
\bibfield{author}{\bibinfo{person}{Immanuel Bayer}, \bibinfo{person}{Xiangnan
  He}, \bibinfo{person}{Bhargav Kanagal}, {and} \bibinfo{person}{Steffen
  Rendle}.} \bibinfo{year}{2017}\natexlab{}.
\newblock \showarticletitle{A generic coordinate descent framework for learning
  from implicit feedback}. In \bibinfo{booktitle}{{\em WWW}}.
  \bibinfo{pages}{1341--1350}.
\newblock


\bibitem[\protect\citeauthoryear{Blattner, Zhang, and Maslov}{Blattner
  et~al\mbox{.}}{2007}]%
        {blattner2007exploring}
\bibfield{author}{\bibinfo{person}{Marcel Blattner}, \bibinfo{person}{Yi-Cheng
  Zhang}, {and} \bibinfo{person}{Sergei Maslov}.}
  \bibinfo{year}{2007}\natexlab{}.
\newblock \showarticletitle{Exploring an opinion network for taste prediction:
  An empirical study}.
\newblock \bibinfo{journal}{{\em Physica A: Statistical Mechanics and its
  Applications\/}} (\bibinfo{year}{2007}), \bibinfo{pages}{753--758}.
\newblock


\bibitem[\protect\citeauthoryear{Cao, Nie, He, Wei, Zhu, and Chua}{Cao
  et~al\mbox{.}}{2017}]%
        {cao2017embedding}
\bibfield{author}{\bibinfo{person}{Da Cao}, \bibinfo{person}{Liqiang Nie},
  \bibinfo{person}{Xiangnan He}, \bibinfo{person}{Xiaochi Wei},
  \bibinfo{person}{Shunzhi Zhu}, {and} \bibinfo{person}{Tat-Seng Chua}.}
  \bibinfo{year}{2017}\natexlab{}.
\newblock \showarticletitle{Embedding Factorization Models for Jointly
  Recommending Items and User Generated Lists}. In \bibinfo{booktitle}{{\em
  SIGIR}}. \bibinfo{pages}{585--594}.
\newblock


\bibitem[\protect\citeauthoryear{Deshpande and Karypis}{Deshpande and
  Karypis}{2004}]%
        {deshpande2004item}
\bibfield{author}{\bibinfo{person}{Mukund Deshpande} {and}
  \bibinfo{person}{George Karypis}.} \bibinfo{year}{2004}\natexlab{}.
\newblock \showarticletitle{Item-based top-n recommendation algorithms}.
\newblock \bibinfo{journal}{{\em TOIS\/}} (\bibinfo{year}{2004}),
  \bibinfo{pages}{143--177}.
\newblock


\bibitem[\protect\citeauthoryear{Devooght, Kourtellis, and Mantrach}{Devooght
  et~al\mbox{.}}{2015}]%
        {devooght2015dynamic}
\bibfield{author}{\bibinfo{person}{Robin Devooght}, \bibinfo{person}{Nicolas
  Kourtellis}, {and} \bibinfo{person}{Amin Mantrach}.}
  \bibinfo{year}{2015}\natexlab{}.
\newblock \showarticletitle{Dynamic matrix factorization with priors on unknown
  values}. In \bibinfo{booktitle}{{\em SIGKDD}}. \bibinfo{pages}{189--198}.
\newblock


\bibitem[\protect\citeauthoryear{Djuric, Wu, Radosavljevic, Grbovic, and
  Bhamidipati}{Djuric et~al\mbox{.}}{2015}]%
        {djuric2015hierarchical}
\bibfield{author}{\bibinfo{person}{Nemanja Djuric}, \bibinfo{person}{Hao Wu},
  \bibinfo{person}{Vladan Radosavljevic}, \bibinfo{person}{Mihajlo Grbovic},
  {and} \bibinfo{person}{Narayan Bhamidipati}.}
  \bibinfo{year}{2015}\natexlab{}.
\newblock \showarticletitle{Hierarchical neural language models for joint
  representation of streaming documents and their content}. In
  \bibinfo{booktitle}{{\em WWW}}. \bibinfo{pages}{248--255}.
\newblock


\bibitem[\protect\citeauthoryear{Gu{\`a}rdia-Sebaoun, Guigue, and
  Gallinari}{Gu{\`a}rdia-Sebaoun et~al\mbox{.}}{2015}]%
        {guardia2015latent}
\bibfield{author}{\bibinfo{person}{Elie Gu{\`a}rdia-Sebaoun},
  \bibinfo{person}{Vincent Guigue}, {and} \bibinfo{person}{Patrick Gallinari}.}
  \bibinfo{year}{2015}\natexlab{}.
\newblock \showarticletitle{Latent trajectory modeling: A light and efficient
  way to introduce time in recommender systems}. In \bibinfo{booktitle}{{\em
  RecSys}}. \bibinfo{pages}{281--284}.
\newblock


\bibitem[\protect\citeauthoryear{He and McAuley}{He and McAuley}{2016}]%
        {he2016ups}
\bibfield{author}{\bibinfo{person}{Ruining He} {and} \bibinfo{person}{Julian
  McAuley}.} \bibinfo{year}{2016}\natexlab{}.
\newblock \showarticletitle{Ups and downs: Modeling the visual evolution of
  fashion trends with one-class collaborative filtering}. In
  \bibinfo{booktitle}{{\em WWW}}. \bibinfo{pages}{507--517}.
\newblock


\bibitem[\protect\citeauthoryear{He, Liao, Zhang, Nie, Hu, and Chua}{He
  et~al\mbox{.}}{2017}]%
        {he2017neural}
\bibfield{author}{\bibinfo{person}{Xiangnan He}, \bibinfo{person}{Lizi Liao},
  \bibinfo{person}{Hanwang Zhang}, \bibinfo{person}{Liqiang Nie},
  \bibinfo{person}{Xia Hu}, {and} \bibinfo{person}{Tat-Seng Chua}.}
  \bibinfo{year}{2017}\natexlab{}.
\newblock \showarticletitle{Neural collaborative filtering}. In
  \bibinfo{booktitle}{{\em WWW}}. \bibinfo{pages}{173--182}.
\newblock


\bibitem[\protect\citeauthoryear{He, Zhang, Kan, and Chua}{He
  et~al\mbox{.}}{2016}]%
        {he2016fast}
\bibfield{author}{\bibinfo{person}{Xiangnan He}, \bibinfo{person}{Hanwang
  Zhang}, \bibinfo{person}{Min-Yen Kan}, {and} \bibinfo{person}{Tat-Seng
  Chua}.} \bibinfo{year}{2016}\natexlab{}.
\newblock \showarticletitle{Fast matrix factorization for online recommendation
  with implicit feedback}. In \bibinfo{booktitle}{{\em SIGIR}}.
  \bibinfo{pages}{549--558}.
\newblock


\bibitem[\protect\citeauthoryear{Hu, Koren, and Volinsky}{Hu
  et~al\mbox{.}}{2008}]%
        {hu2008collaborative}
\bibfield{author}{\bibinfo{person}{Yifan Hu}, \bibinfo{person}{Yehuda Koren},
  {and} \bibinfo{person}{Chris Volinsky}.} \bibinfo{year}{2008}\natexlab{}.
\newblock \showarticletitle{Collaborative filtering for implicit feedback
  datasets}. In \bibinfo{booktitle}{{\em ICDM}}. \bibinfo{pages}{263--272}.
\newblock


\bibitem[\protect\citeauthoryear{Koren}{Koren}{2008}]%
        {koren2008factorization}
\bibfield{author}{\bibinfo{person}{Yehuda Koren}.}
  \bibinfo{year}{2008}\natexlab{}.
\newblock \showarticletitle{Factorization meets the neighborhood: a
  multifaceted collaborative filtering model}. In \bibinfo{booktitle}{{\em
  SIGKDD}}. \bibinfo{pages}{426--434}.
\newblock


\bibitem[\protect\citeauthoryear{Koren}{Koren}{2009}]%
        {Koren09}
\bibfield{author}{\bibinfo{person}{Yehuda Koren}.}
  \bibinfo{year}{2009}\natexlab{}.
\newblock \showarticletitle{Collaborative filtering with temporal dynamics}. In
  \bibinfo{booktitle}{{\em SIGKDD}}. \bibinfo{pages}{447--456}.
\newblock


\bibitem[\protect\citeauthoryear{Koren, Bell, and Volinsky}{Koren
  et~al\mbox{.}}{2009}]%
        {koren2009matrix}
\bibfield{author}{\bibinfo{person}{Yehuda Koren}, \bibinfo{person}{Robert
  Bell}, {and} \bibinfo{person}{Chris Volinsky}.}
  \bibinfo{year}{2009}\natexlab{}.
\newblock \showarticletitle{Matrix factorization techniques for recommender
  systems}.
\newblock \bibinfo{journal}{{\em Computer\/}} (\bibinfo{year}{2009}).
\newblock


\bibitem[\protect\citeauthoryear{Le and Mikolov}{Le and Mikolov}{2014}]%
        {le2014distributed}
\bibfield{author}{\bibinfo{person}{Quoc Le} {and} \bibinfo{person}{Tomas
  Mikolov}.} \bibinfo{year}{2014}\natexlab{}.
\newblock \showarticletitle{Distributed representations of sentences and
  documents}. In \bibinfo{booktitle}{{\em ICML}}. \bibinfo{pages}{1188--1196}.
\newblock


\bibitem[\protect\citeauthoryear{Levy and Goldberg}{Levy and Goldberg}{2014}]%
        {levy2014neural}
\bibfield{author}{\bibinfo{person}{Omer Levy} {and} \bibinfo{person}{Yoav
  Goldberg}.} \bibinfo{year}{2014}\natexlab{}.
\newblock \showarticletitle{Neural word embedding as implicit matrix
  factorization}. In \bibinfo{booktitle}{{\em NIPS}}.
  \bibinfo{pages}{2177--2185}.
\newblock


\bibitem[\protect\citeauthoryear{Liang, Altosaar, Charlin, and Blei}{Liang
  et~al\mbox{.}}{2016a}]%
        {liang2016factorization}
\bibfield{author}{\bibinfo{person}{Dawen Liang}, \bibinfo{person}{Jaan
  Altosaar}, \bibinfo{person}{Laurent Charlin}, {and} \bibinfo{person}{David~M
  Blei}.} \bibinfo{year}{2016}\natexlab{a}.
\newblock \showarticletitle{Factorization meets the item embedding:
  Regularizing matrix factorization with item co-occurrence}. In
  \bibinfo{booktitle}{{\em RecSys}}. \bibinfo{pages}{59--66}.
\newblock


\bibitem[\protect\citeauthoryear{Liang, Charlin, McInerney, and Blei}{Liang
  et~al\mbox{.}}{2016b}]%
        {liang2016modeling}
\bibfield{author}{\bibinfo{person}{Dawen Liang}, \bibinfo{person}{Laurent
  Charlin}, \bibinfo{person}{James McInerney}, {and} \bibinfo{person}{David~M
  Blei}.} \bibinfo{year}{2016}\natexlab{b}.
\newblock \showarticletitle{Modeling user exposure in recommendation}. In
  \bibinfo{booktitle}{{\em WWW}}. \bibinfo{pages}{951--961}.
\newblock


\bibitem[\protect\citeauthoryear{Liu, Lee, Yu, and Li}{Liu
  et~al\mbox{.}}{2002}]%
        {liu2002partially}
\bibfield{author}{\bibinfo{person}{Bing Liu}, \bibinfo{person}{Wee~Sun Lee},
  \bibinfo{person}{Philip~S Yu}, {and} \bibinfo{person}{Xiaoli Li}.}
  \bibinfo{year}{2002}\natexlab{}.
\newblock \showarticletitle{Partially supervised classification of text
  documents}. In \bibinfo{booktitle}{{\em ICML}}. \bibinfo{pages}{387--394}.
\newblock


\bibitem[\protect\citeauthoryear{McAuley and Leskovec}{McAuley and
  Leskovec}{2013}]%
        {mcauley2013hidden}
\bibfield{author}{\bibinfo{person}{Julian McAuley} {and} \bibinfo{person}{Jure
  Leskovec}.} \bibinfo{year}{2013}\natexlab{}.
\newblock \showarticletitle{Hidden factors and hidden topics: understanding
  rating dimensions with review text}. In \bibinfo{booktitle}{{\em RecSys}}.
  \bibinfo{pages}{165--172}.
\newblock


\bibitem[\protect\citeauthoryear{Mikolov, Sutskever, Chen, Corrado, and
  Dean}{Mikolov et~al\mbox{.}}{2013}]%
        {mikolov2013distributed}
\bibfield{author}{\bibinfo{person}{Tomas Mikolov}, \bibinfo{person}{Ilya
  Sutskever}, \bibinfo{person}{Kai Chen}, \bibinfo{person}{Greg~S Corrado},
  {and} \bibinfo{person}{Jeff Dean}.} \bibinfo{year}{2013}\natexlab{}.
\newblock \showarticletitle{Distributed representations of words and phrases
  and their compositionality}. In \bibinfo{booktitle}{{\em NIPS}}.
  \bibinfo{pages}{3111--3119}.
\newblock


\bibitem[\protect\citeauthoryear{Pan, Zhou, Cao, Liu, Lukose, Scholz, and
  Yang}{Pan et~al\mbox{.}}{2008}]%
        {pan2008one}
\bibfield{author}{\bibinfo{person}{Rong Pan}, \bibinfo{person}{Yunhong Zhou},
  \bibinfo{person}{Bin Cao}, \bibinfo{person}{Nathan~N Liu},
  \bibinfo{person}{Rajan Lukose}, \bibinfo{person}{Martin Scholz}, {and}
  \bibinfo{person}{Qiang Yang}.} \bibinfo{year}{2008}\natexlab{}.
\newblock \showarticletitle{One-class collaborative filtering}. In
  \bibinfo{booktitle}{{\em ICDM}}. \bibinfo{pages}{502--511}.
\newblock


\bibitem[\protect\citeauthoryear{Pennington, Socher, and Manning}{Pennington
  et~al\mbox{.}}{2014}]%
        {pennington2014glove}
\bibfield{author}{\bibinfo{person}{Jeffrey Pennington},
  \bibinfo{person}{Richard Socher}, {and} \bibinfo{person}{Christopher~D
  Manning}.} \bibinfo{year}{2014}\natexlab{}.
\newblock \showarticletitle{Glove: Global vectors for word representation}. In
  \bibinfo{booktitle}{{\em EMNLP}}. \bibinfo{pages}{1532--1543}.
\newblock


\bibitem[\protect\citeauthoryear{Pil{\'a}szy, Zibriczky, and Tikk}{Pil{\'a}szy
  et~al\mbox{.}}{2010}]%
        {pilaszy2010fast}
\bibfield{author}{\bibinfo{person}{Istv{\'a}n Pil{\'a}szy},
  \bibinfo{person}{D{\'a}vid Zibriczky}, {and} \bibinfo{person}{Domonkos
  Tikk}.} \bibinfo{year}{2010}\natexlab{}.
\newblock \showarticletitle{Fast als-based matrix factorization for explicit
  and implicit feedback datasets}. In \bibinfo{booktitle}{{\em RecSys}}.
  \bibinfo{pages}{71--78}.
\newblock


\bibitem[\protect\citeauthoryear{Rendle, Freudenthaler, Gantner, and
  Schmidt-Thieme}{Rendle et~al\mbox{.}}{2009}]%
        {rendle2009bpr}
\bibfield{author}{\bibinfo{person}{Steffen Rendle}, \bibinfo{person}{Christoph
  Freudenthaler}, \bibinfo{person}{Zeno Gantner}, {and} \bibinfo{person}{Lars
  Schmidt-Thieme}.} \bibinfo{year}{2009}\natexlab{}.
\newblock \showarticletitle{BPR: Bayesian personalized ranking from implicit
  feedback}. In \bibinfo{booktitle}{{\em UAI}}. \bibinfo{pages}{452--461}.
\newblock


\bibitem[\protect\citeauthoryear{Resnick, Iacovou, Suchak, Bergstrom, and
  Riedl}{Resnick et~al\mbox{.}}{1994}]%
        {resnick1994grouplens}
\bibfield{author}{\bibinfo{person}{Paul Resnick}, \bibinfo{person}{Neophytos
  Iacovou}, \bibinfo{person}{Mitesh Suchak}, \bibinfo{person}{Peter Bergstrom},
  {and} \bibinfo{person}{John Riedl}.} \bibinfo{year}{1994}\natexlab{}.
\newblock \showarticletitle{GroupLens: an open architecture for collaborative
  filtering of netnews}. In \bibinfo{booktitle}{{\em CSCW}}.
  \bibinfo{pages}{175--186}.
\newblock


\bibitem[\protect\citeauthoryear{Salakhutdinov, Mnih, and Hinton}{Salakhutdinov
  et~al\mbox{.}}{2007}]%
        {salakhutdinov2007restricted}
\bibfield{author}{\bibinfo{person}{Ruslan Salakhutdinov},
  \bibinfo{person}{Andriy Mnih}, {and} \bibinfo{person}{Geoffrey Hinton}.}
  \bibinfo{year}{2007}\natexlab{}.
\newblock \showarticletitle{Restricted Boltzmann machines for collaborative
  filtering}. In \bibinfo{booktitle}{{\em ICML}}. \bibinfo{pages}{791--798}.
\newblock


\bibitem[\protect\citeauthoryear{Sarwar, Karypis, Konstan, and Riedl}{Sarwar
  et~al\mbox{.}}{2001}]%
        {sarwar2001item}
\bibfield{author}{\bibinfo{person}{Badrul Sarwar}, \bibinfo{person}{George
  Karypis}, \bibinfo{person}{Joseph Konstan}, {and} \bibinfo{person}{John
  Riedl}.} \bibinfo{year}{2001}\natexlab{}.
\newblock \showarticletitle{Item-based collaborative filtering recommendation
  algorithms}. In \bibinfo{booktitle}{{\em WWW}}. \bibinfo{pages}{285--295}.
\newblock


\bibitem[\protect\citeauthoryear{Steck}{Steck}{2010}]%
        {steck2010training}
\bibfield{author}{\bibinfo{person}{Harald Steck}.}
  \bibinfo{year}{2010}\natexlab{}.
\newblock \showarticletitle{Training and testing of recommender systems on data
  missing not at random}. In \bibinfo{booktitle}{{\em SIGKDD}}.
  \bibinfo{pages}{713--722}.
\newblock


\bibitem[\protect\citeauthoryear{Su and Khoshgoftaar}{Su and
  Khoshgoftaar}{2009}]%
        {SuK09}
\bibfield{author}{\bibinfo{person}{Xiaoyuan Su} {and} \bibinfo{person}{Taghi~M.
  Khoshgoftaar}.} \bibinfo{year}{2009}\natexlab{}.
\newblock \showarticletitle{A Survey of Collaborative Filtering Techniques}.
\newblock \bibinfo{journal}{{\em Adv. Artificial Intellegence\/}}
  (\bibinfo{year}{2009}).
\newblock


\bibitem[\protect\citeauthoryear{Volkovs and Yu}{Volkovs and Yu}{2015}]%
        {volkovs2015effective}
\bibfield{author}{\bibinfo{person}{Maksims Volkovs} {and}
  \bibinfo{person}{Guang~Wei Yu}.} \bibinfo{year}{2015}\natexlab{}.
\newblock \showarticletitle{Effective latent models for binary feedback in
  recommender systems}. In \bibinfo{booktitle}{{\em SIGIR}}.
  \bibinfo{pages}{313--322}.
\newblock


\bibitem[\protect\citeauthoryear{Wang and Blei}{Wang and Blei}{2011}]%
        {wang2011collaborative}
\bibfield{author}{\bibinfo{person}{Chong Wang} {and} \bibinfo{person}{David~M
  Blei}.} \bibinfo{year}{2011}\natexlab{}.
\newblock \showarticletitle{Collaborative topic modeling for recommending
  scientific articles}. In \bibinfo{booktitle}{{\em SIGKDD}}.
  \bibinfo{pages}{448--456}.
\newblock


\bibitem[\protect\citeauthoryear{Wang, Li, Wang, and Zeng}{Wang
  et~al\mbox{.}}{2017}]%
        {wang2017incorporating}
\bibfield{author}{\bibinfo{person}{Can Wang}, \bibinfo{person}{Qiudan Li},
  \bibinfo{person}{Lei Wang}, {and} \bibinfo{person}{Daniel~Dajun Zeng}.}
  \bibinfo{year}{2017}\natexlab{}.
\newblock \showarticletitle{Incorporating message embedding into co-factor
  matrix factorization for retweeting prediction}. In \bibinfo{booktitle}{{\em
  IJCNN}}. \bibinfo{pages}{1265--1272}.
\newblock


\bibitem[\protect\citeauthoryear{Xue, Dai, Zhang, Huang, and Chen}{Xue
  et~al\mbox{.}}{2017}]%
        {xue2017deep}
\bibfield{author}{\bibinfo{person}{Hong-Jian Xue}, \bibinfo{person}{Xin-Yu
  Dai}, \bibinfo{person}{Jianbing Zhang}, \bibinfo{person}{Shujian Huang},
  {and} \bibinfo{person}{Jiajun Chen}.} \bibinfo{year}{2017}\natexlab{}.
\newblock \showarticletitle{Deep matrix factorization models for recommender
  systems}.
\newblock \bibinfo{journal}{{\em static. ijcai. org\/}} (\bibinfo{year}{2017}).
\newblock


\bibitem[\protect\citeauthoryear{Yu, Hsieh, Si, and Dhillon}{Yu
  et~al\mbox{.}}{2014}]%
        {yu2014parallel}
\bibfield{author}{\bibinfo{person}{Hsiang-Fu Yu}, \bibinfo{person}{Cho-Jui
  Hsieh}, \bibinfo{person}{Si Si}, {and} \bibinfo{person}{Inderjit~S Dhillon}.}
  \bibinfo{year}{2014}\natexlab{}.
\newblock \showarticletitle{Parallel matrix factorization for recommender
  systems}.
\newblock \bibinfo{journal}{{\em Knowledge and Information Systems\/}}
  (\bibinfo{year}{2014}), \bibinfo{pages}{793--819}.
\newblock


\bibitem[\protect\citeauthoryear{Zhang, Chen, Wang, and Yu}{Zhang
  et~al\mbox{.}}{2013}]%
        {zhang2013optimizing}
\bibfield{author}{\bibinfo{person}{Weinan Zhang}, \bibinfo{person}{Tianqi
  Chen}, \bibinfo{person}{Jun Wang}, {and} \bibinfo{person}{Yong Yu}.}
  \bibinfo{year}{2013}\natexlab{}.
\newblock \showarticletitle{Optimizing top-n collaborative filtering via
  dynamic negative item sampling}. In \bibinfo{booktitle}{{\em SIGIR}}.
  \bibinfo{pages}{785--788}.
\newblock


\bibitem[\protect\citeauthoryear{Zhou, Wilkinson, Schreiber, and Pan}{Zhou
  et~al\mbox{.}}{2008}]%
        {zhou2008large}
\bibfield{author}{\bibinfo{person}{Yunhong Zhou}, \bibinfo{person}{Dennis
  Wilkinson}, \bibinfo{person}{Robert Schreiber}, {and} \bibinfo{person}{Rong
  Pan}.} \bibinfo{year}{2008}\natexlab{}.
\newblock \showarticletitle{Large-scale parallel collaborative filtering for
  the netflix prize}. In \bibinfo{booktitle}{{\em International Conference on
  Algorithmic Applications in Management}}. \bibinfo{pages}{337--348}.
\newblock


\end{thebibliography}
			
		\end{document}